# The co-evolutionary relationship between digitalization and organizational agility: Ongoing debates, theoretical developments and future research perspectives

Highlights

- The first systematic literature review on the digitalization-OA relationships.
- Uses the dynamic capabilities' lenses to achieve a holistic understanding of the current and future research pathways.
- Paves the way for a new corpus of studies investigating the co-evolutionary interconnections between digitalization and OA.
- Stimulates a new managerial two-way thinking approach to the digitalization-OA relationships.
- Three main thematic clusters are discovered and thirteen new research and managerial perspectives are developed.

# The co-evolutionary relationship between digitalization and organizational agility: Ongoing debates, theoretical developments andfuture research perspectives

**Francesco Ciampi***
Full Professor
Department of Economics and Management, University of Florence,
Via delle Pandette 9 - Block D6, 3rd Floor, 50127 Florence (Italy)
E-mail: francesco.ciampi@unifi.it

**Monica Faraoni**
Associate Professor
Department of Economics and Management, University of Florence,
Via delle Pandette 9 - Block D6, 3rd Floor, 50127 Florence (Italy)
E-mail: monica.faraoni@unifi.it

**Jacopo Ballerini**
Lecturer of Innovation Management and Digital
TransformationDepartment of Management, University of
Turin,
Corso Unione Sovietica 218/bis, 10134 Turin
(Italy)E-mail: jacopo.ballerini@unito.it

**Francesco Meli**
External Collaborator of the Department
Department of Economics and Management, University of Florence,
Via delle Pandette 9 - Block D6, 3rd Floor, 50127 Florence (Italy)
francesco.meli654@gmail.com

**Declarations of interest: none**
**This research did not receive any specific grant from funding agencies in the public,commercial, or not-for-profit sectors.**

**\* Corresponding Author:**
Francesco
CiampiFull
Professor
Department of Economics and Management, University of
Florence,Via delle Pandette 9 - Block D6, 3rd Floor, 50127
Florence (Italy)
E-mail:
francesco.ciampi@unifi.it
Mobile +39 335 5893497

# The co-evolutionary relationship between digitalization and organizational agility: Ongoing debates, theoretical developments and future research perspectives

**Abstract**

This study is the first to provide a systematic review of the literature focused on the relationship between digitalization and organizational agility (OA). It applies the bibliographic coupling method to 171 peer-reviewed contributions published by 30 June 2021. It uses the digitalization perspective to investigate the enablers, barriers and benefits of processes aimed at providing firms with the agility required to effectively face increasingly turbulent environments. Three different, though interconnected, thematic clusters are discovered and analysed, respectively focusing on big-data analytic capabilities as crucial drivers of OA, the relationship between digitalization and agility at a supply chain level, and the role of information technology capabilities in improving OA. By adopting a dynamic capabilities perspective, this study overcomes the traditional view, which mainly considers digital capabilities enablers of OA, rather than as possible outcomes. Our findings reveal that, in addition to being complex, the relationship between digitalization and OA has a bidirectional character. This study also identifies extant research gaps and develops 13 original research propositions on possible future research pathways and new managerial solutions.

**Paper type**: Literature review

## 1. Introduction

The global economy is evolving rapidly, and most marketplaces are highly turbulent due to the ever-changing market dynamics, the constantly evolving needs of consumers and workers, and the continuous flow of new digital technologies. The digital revolution is disrupting global competitiveness and paving the way for strong improvements in the efficiency, productivity, fluidity, flexibility and effectiveness of millions of companies' business processes (Schwab and Zahidi, 2020). The pandemic that shook the world at the beginning of 2020 emphasized this turbulence (Accenture, 2020), making it all the more relevant for firms to respond agilely to internal and external changes, including those concerning the increasingly dynamic and turbulent digital landscape.

Organizational agility (OA) is the ability to quickly detect and analyse opportunities and threats, even in their latent state, and to respond to them effectively by adopting required changes and actions (Barlette and Baillette, 2020; Felipe et al., 2020). In our era of growing complexity and uncertainty, agility is a crucial capability (Overby et al., 2006; Teece, 2016) that allows firms to respond appropriately to continuous environmental changes, seize emerging business opportunities, defend and strengthen their competitive advantages (Inman et al., 2011; Vickery et al., 2010) as well as enhance their overall financial performance (Hatzijordanou et al., 2019). In fact, agile firms generate higher revenues and profits than non-agile organizations do (Glenn, 2009; Wang et al., 2013).

Agility is also crucial for exploiting ever-evolving digital technologies effectively (Brenner, 2018). Scholars have estimated that, firms can increase their profit margins by more than 80% by leveraging digitalization (Škare and Soriano, 2021). A large part of digitalization's potential value creation effect on company performance depends on the positive impact that new digital technologies exert on OA. Consider, for example, how big data analytics (BDA) and the Internet of Things (IoT) currently allow the effective collection, structure, management, sharing and interpretation of huge volumes of information (Prescott, 2014; McAfee and Brynjolfsson, 2012). In addition, BDA and IoT provide insights into real-time effective marketing investigations (Ahn, 2020; von Alberti-Alhtaybat et al., 2019). Furthermore, blockchain technologies ensure significant and scalable processing power and high levels of accuracy and security (Nandi et al., 2020; Saberi et al., 2019; Sheel and Nath, 2019), while simultaneously allowing the elimination of many wasteful and costly control tasks (Gunasekaran et al., 2019; Nandi et al., 2020; Rane and Narvel, 2019). Moreover, though managers should cautiously and accurately design and implement precise schedules and guidelines aimed at minimizing their adverse effects on employee creativity and wellbeing (Luqman et al., 2021), enterprise social media facilitate day-to-day workflow agility (Huang et al., 2015; Kane, 2015; Pitafi et al., 2020) and may also help prevent the cyber-slacking effect, i.e., employees' personal use of the Internet for non-work purposes (Nusrat et al., 2021). Finally, enterprise social media facilitate day-to-day workflow agility (Huang et al., 2015; Kane, 2015; Pitafi et al., 2020), the IoT reduces machine maintenance times, efforts and costs (Dijkman et al., 2015; Rane and Narvel, 2019), while digital cloud-based technologies and big data enhance IT infrastructure's flexibility and the integration between functional departments (Fosso Wamba et al., 2020; Liu et al., 2018).

The literature has analysed both OA and digital capabilities, using quite complex constructs and different approaches (Škare and Soriano, 2021). A relevant stream of literature investigates digital capabilities' impact on organizational agility. These capabilities are regarded as those needed to timely and successfully design and implement digital transformation (Chakravarty et al., 2013). This literature finds that strategically aligning and integrating digital infrastructure and capabilities cross-functionally, are fundamental for driving agility, since they allow to identify market opportunities, acquire the resources needed to seize these opportunities (Tallon and Pinsonneault, 2011), manage information exchanges effectively and overcome the barriers that normally impede organizational agility (Pinsonneault and Kraemer, 2002). IT-based decision systems and data warehouses help firms monitor data in real time, recognize new trends, and subsequently change their strategy (Wixom and Watson, 2001). According to this literature, firms need to constantly explore all digital technologies' emerging applications and continuously enhance and leverage their digital capabilities in order to gain the agility required to effectively face an increasingly volatile and challenging future (e.g., Akhtar et al., 2019; Barlette and Baillette, 2020; Kozarkiewicz, 2020; Malekifar et al., 2014; Panda and Rath, 2017, 2016).

However, by their very nature, agile organizations are simultaneously well positioned to respond rapidly and effectively to internal and external changes, including those concerning the increasingly dynamic and turbulent digital landscape. Consequently, scholars also view digital capabilities as an outcome of an organization's agility (Nwankpa and Merhout, 2020). Further, scholars view an organization's agility as a crucial firm lever for crafting an effective digital company mind-set (Warner and Wäger, 2019) and for successfully exploiting emerging digital technologies such as BDA, cloud computing, blockchain, and IoT (Brenner, 2018; Vial, 2019).

We found literature reviews on overall organizational agility (Walter, 2020), workforce agility (Muduli, 2013; Al-Kasasbeh et al., 2016), agility in the manufacturing industry (Gunasekaran, 1999; Potdar et al., 2017; Yusuf et al., 1999), supply chain agility (Gligor et al., 2019; Gligor and Holcomb, 2012), agility enablers (Marhraoui and El Manouar, 2017) and IT for organizational agility (Tallon et al., 2019). Other reviews covered digitalization's effect on organizations (Kuusisto, 2017), the digitalization phenomenon analysed by means of different approaches (Nadkarni and Prügl, 2021; Verhoef et al., 2021; Vial, 2019), the digitalization of the supply chain (Shashi et al., 2020), digital innovation in knowledge management systems (Di Vaio et al., 2021), digital servitization in manufacturing (Paschou et al., 2020), the digitalization of SMEs (Isensee et al., 2020), and the effects of specific types of digital technologies, such as IoT (Del Giudice, 2016), big data (Khanra et al., 2020a; Talwar et al., 2021) and blockchain (Wang et al., 2019).

Digital technologies and agility are of crucial importance for firms in respect of allowing them to effectively face the challenges of a volatile, uncertain, complex and ambiguous (VUCA) environment. Nevertheless, despite the intrinsic complexity of the two constructs and their co-evolutionary and articulated interconnections, the literature does not, to the best of the authors' knowledge, offer a holistic and comprehensive understanding of the existing body of knowledge concerning the relationships between digitalization and agility, the research gaps and understudied topics to be addressed, the emergent issues that have been under investigated and the most promising future research avenues.

With this study, we aim to fill this gap by attempting to answer four specific research questions (**RQs**).

**RQ1.** Which are the key research themes of the existing literature on the relationships between digitalization and organizational agility?

In this connection, our study aims to paint an up-to-date picture of the literature by developing a systematic thematic map on which we will base the answers to the subsequent research questions.

**RQ2.** Which are this literature's main research gaps and limitations?

**RQ3.** Which are the field's more promising future research avenues?

With regard to RQs 2 and 3, our study aims to identify extant research gaps and develop a viable agenda comprising a set of original propositions on new research avenues and managerial issues for future exploration. Although mainly focused on the relationships between agility and digital capabilities, approaches and culture, most of the propositions developed in this study could also stimulate further research into the digitalization and agility's effects on corporate strategy and innovation (Ciampi et al., 2020).

**RQ4.** Is the relationship between digitalization and agility, a one-way or a two-way relationship?

By exploring a two-way relationship perspective to analyse the interconnections between agility and digitalization, this study aims to overcome the traditional one-direction thinking, according to which digital transformation fosters OA and not vice versa (e.g., Ahn, 2020; Akhtar et al., 2019; Panda and Rath, 2017, 2016). In so doing, our study paves the way for a new corpus of studies investigating and exploiting the mutual and co-evolutionary interconnections between digitalization and OA. In addition, the study encourages a new managerial approach of two-way thinking, which allows managers to effectively design agility and digital capabilities' balanced development and to successfully manage their complex interconnections.

This study contributes to the existing literature by identifying and analysing three different, though interconnected, thematic clusters, namely focusing on big-data analytic capabilities as crucial drivers of OA, the relationship between digitalization and agility at a supply chain level, and the role of information technology capabilities in improving OA. Our findings reveal that effectively leveraging digital technologies to enhance OA requires the digitalization processes to also be agile and flexible, as these allow the continuous acquisition of new digital capabilities and the latter's effective integration with the existing ones. This study also develops and presents a possible future research agenda based on the four new research pathways that we develop for each of the mentioned thematic clusters. In addition, the study also develops a final research pathway related to the bidirectional nature of the relationship between digital capabilities and agility capabilities and the importance of the balanced development of both sets of capabilities for a firm's competitiveness and performance. We also find that the portfolio of agility capabilities that a firm possesses must evolve progressively during the various phases of its digital development (Vokurka and Fliedner, 1998) and propose a framework aimed at giving insights into the different agility capabilities that are critical in the digital transformation lifecycle's various phases.

The next section presents our review's theoretical background, while section 3 presents the methodology we adopted. We present the results of the bibliometric and the visualization of similarities (VOS) analyses in section 4, followed by a systematic review of the literature in section 5. In section 6, we discuss our findings by proposing a future research agenda based on 13 original research propositions (section 6.1) and synthesising our study's main theoretical contributions (section 6.2), managerial implications (section 6.3), and limitations (section 6.4). A concluding section briefly summarizes the study's key findings.

## 2. Theoretical background

OA is a firm's ability to accelerate decision making, sense marketing and environmental changes quickly and flexibly as well as to exploit emerging opportunities effectively and to open new avenues of competitive advantage (Overby et al., 2006; Sambamurthy et al., 2003; Tallon and Pinsonneault, 2011). Starting with studies on the field of flexible manufacturing systems (Sarker et al., 2009; White et al., 2005), management scholars have investigated OA from diverse perspectives. Different types of OA have been identified and analysed, from workforce agility (Crocitto and Youssef, 2003; Patil and Suresh, 2019; Sommer, 2019) to logistic agility (Cao and Dowlatshahi, 2005; Zielske and Held, 2020), from marketing agility (Osei et al., 2018; Zhou et al., 2019) to strategic agility (Cunha et al., 2020; Doz, 2020), and supply chain agility (Chen, 2019; Christopher and Ryals, 2014). Adopting a dynamic capabilities (DCs) theory perspective, scholars view OA as responsiveness, competency, flexibility, speed (Lin et al., 2006; Zhang and Sharifi, 2000), proactiveness, radicalness, and adaptiveness (Lee et al., 2015), or, more simply, as the capability to sense change and respond it (Overby et al., 2006).

Achieving high levels of agility is no easy task. The literature has identified several OA antecedents, ranging from technological ones, such as Information Technology (IT) and its role in improving the company and the supply chain capabilities regarding adapting to and anticipating environmental changes (Sambamurthy et al., 2003; Swafford et al., 2006), to behavioural antecedents, such as management influential leaderships styles and risk-taking mindsets as well as company innovation capacity and culture (Breu et al., 2002; Tallon, 2008); and also ranging from organizational antecedents, such as those concerning strategic orientations and business models, to environmental antecedents, such as those concerning environmental uncertainty and dynamism (Tallon, et al., 2019).

Agility is a key capability also with regard to addressing the digital revolution currently disrupting global competitiveness timely and effectively (Schwab and Zahidi, 2020). In today's world, to survive and prosper companies must face this revolution appropriately (Chan et al., 2019; Lucas and Goh, 2009) by continuously evolving their business models, structures and processes (Hess et al., 2016), as well as their culture and approach to strategic change and collaboration (Warner and Wäger, 2019). At the same time, developing adequate digital skills and capabilities (Li et al., 2018), while effectively integrating digital technologies and business processes simultaneously (Liu et al., 2018; Vial, 2019) are currently required steps for firms endeavouring to cope with the challenges of an increasingly volatile, uncertain, complex and ambiguous competitive world (Millar et al., 2018).

The current digital revolution is a disruptive innovation (Karimi and Walter, 2015; Vial, 2019) based on a new generation of information technologies, such as data analytics, IoT and social media (Fitzgerald et al., 2014). This revolution disrupts consumer behaviour and expectations by allowing clients to assume a proactive role in their dialogue with organizations and in the co-creation processes of goods and services; it also disrupts business models (Ciampi et al., 2021) and the competitive landscape by moving competition from physical to virtual products and environments (Günther et al., 2017; Vial, 2019). All of the latter require new capabilities for managing digital instruments and technologies (Khin and Ho, 2018; Kane et al., 2015).

The digital revolution's impact on company value chains varies in keeping with the digital evolution process's various phases. In the digitization stage, during which the analogical to digital information conversion takes place (Verhoef et al., 2021), company value creation processes are not significantly affected. In the digitalization phase, during which IT and digital technologies act as key enablers of business activities' transformation and optimization processes, such as communication, manufacturing or customer relationship management, a relevant restructuration of the operative value creation activities occurs. Finally, in the digital transformation phase, the usual outcomes are new business models and new ways of interacting with the competitive environment, which deeply impact the value creation system (Seetharaman, 2020).

The relationship between agility and digitalization are complex, articulated and partly controversial. Scholars have identified IT and digital capabilities as crucial enablers of several aspects of organizational agility, such as operational agility (Liu et al., 2018; Sambamurthy et al., 2003; Vagnoni and Khoddami, 2016), business process agility (Panda and Rath, 2016), market exploiting agility (Lu and K. (Ram) Ramamurthy, 2011; Mao et al., 2015; Melián-Alzola et al., 2020), and sensing and responding agility (Panda and Rath, 2017). At the same time, agility is a natural lever for firms to successfully exploit the ever-evolving digital technologies (Brenner, 2018; Vial, 2019) and develop an effective company digital culture (Warner and Wäger, 2019). Consequently, digital capabilities can also be viewed as an outcome of an organization's agility (Nwankpa and Merhout, 2020).

Besides being tricky and challenging, the interactions between digital capabilities and OA are also crucial drivers of organizational innovation (Cai et al., 2019; Cepeda and Arias-Pérez, 2019) and performance (Liu et al., 2013; Martínez-Caro et al., 2020; Sambamurthy et al., 2003), even at a supply chain level (Alzoubi and Yanamandra, 2020; Chen, 2019; Swafford et al., 2008).

This study investigates digitalization and agility by adopting the dynamic capabilities (DCs) perspective. Teece et al (1997) define DCs as "the firm's ability to integrate, build, and reconfigure internal and external competences to address rapidly changing environments". These capabilities are crucial for effectively managing the current VUCA era challenges that we are experiencing (Teece et al., 2016). Both OA and digital capabilities are important for counteracting environmental uncertainty and dynamism, and have an inherently dynamic nature (Millar et al., 2018; Teece et al., 2016). DCs differ from operational routines in that they refer to the firm's ability to reconfigure and integrate their resources to effectively and timely sense and respond to environmental changes (Ayabakan et al., 2017; Teece, 2007). Agility embodies all these characteristics and can also be leveraged to reconfigure and create other capabilities; consequently, we can view OA as a subset of DCs (Ghasemaghaei et al., 2017) that enables companies to confront environmental evolution and demand changes effectively (Overby et al., 2006), and to effectively redirect resources towards higher-yield value creation, protection and capturing activities in response to internal and external circumstances' evolution (Teece, 2016).

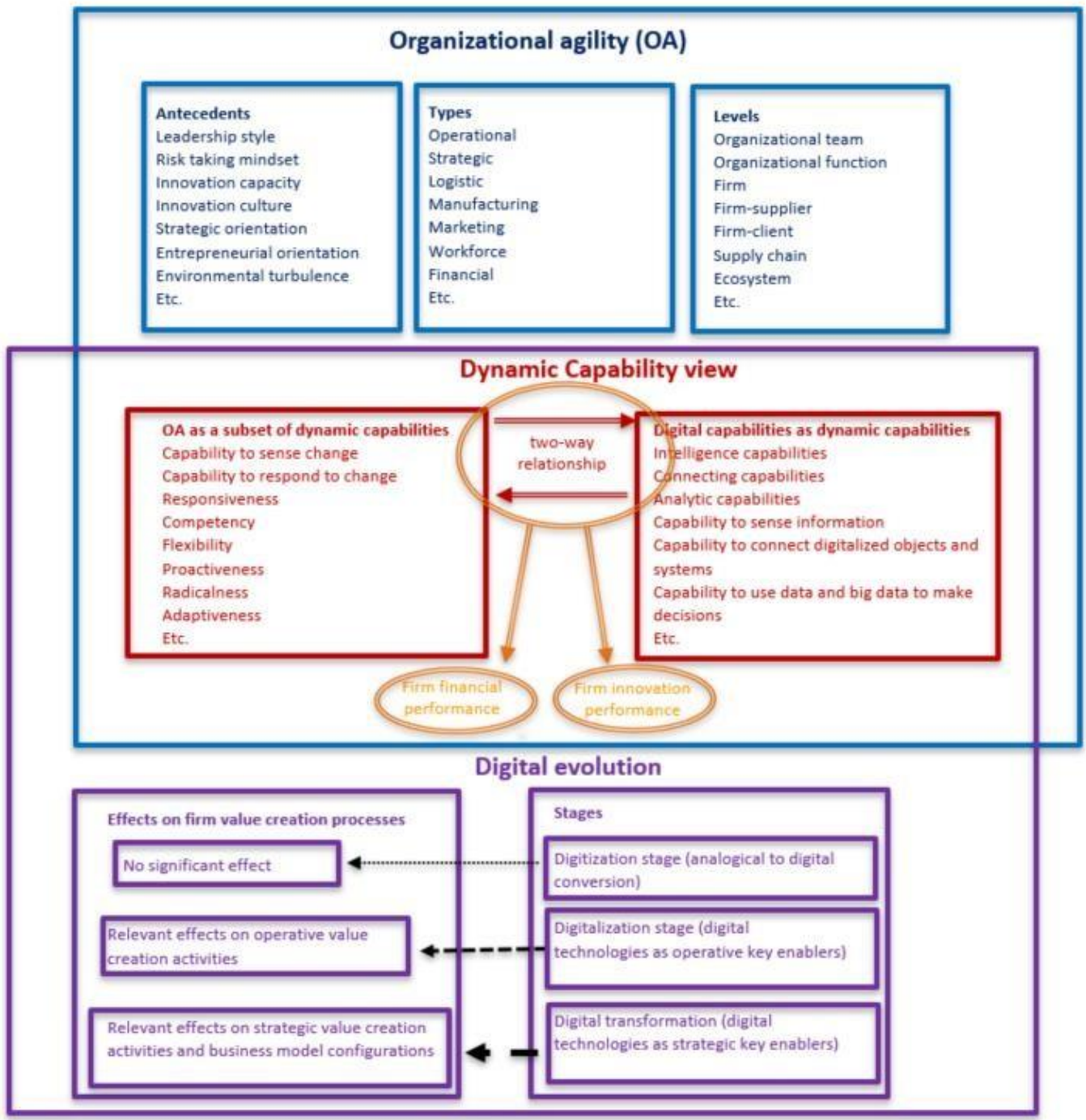

**Figure 1.** Theoretical background: A concise schematization

At the same time, effectively detecting, designing, adopting and exploiting the new and ever evolving instruments and technologies that the digital revolution makes available, require firms to develop an adequate set of digital capabilities enabling companies to coherently transform and evolve their operational processes, business models and customer experiences (Westerman et al., 2011), all of which are another crucial emerging dynamic capability category (Warner and Wäger, 2019). Digital capabilities, i.e. the capabilities needed to timely and successfully design and implement digital transformation, are based on two building blocks: a solid information management ability and an effective IT infrastructure (Levallet and Chan, 2018).Thee capabilities are based on intelligence capabilities, connecting capabilities, and analytic capabilities. Analytic capabilities are the ability to configure hardware components to sense information, connecting capabilities refer to the ability to connect digitalized products through wireless communication systems and analytic capabilities are the ability to use the available data to make business decisions (Lenka et al., 2017). Based on these considerations, among the two main perspectives of viewing OA and digitalization as either a paradigm (e.g., Zhang and Sharifi, 2000) or a capability (e.g., Bessant et al., 2001), this study adopts the DCs view.

## 3. Methodology

We undertook our systematic literature review by adapting the innovative protocols that Behera et al. (2019) recently developed on the basis of a comprehensive and robust analysis of the criteria used in previous reviews. Besides being novel, these protocols have recently proved to be particularly effective for accountable, coherent and replicable systematic literature reviews (Kaur et al., 2020; Khan et al., 2021; Tandon et al., 2020). They are therefore an effective methodological tool for addressing the study's research questions.

In line with the literature (Bresciani et al., 2021; Kaur et al., 2020; Kraus et al., 2020; Tranfield et al., 2003), we followed a reproducible and rigorous process when selecting the sample of papers to analyse. Our review development comprised three principal stages: (i) planning the review, (ii) undertaking the review, and (iii) developing the review (see Figure 2).

In the first stage, we outlined the research objectives, identified the inclusion and exclusion criteria and selected an appropriate database. The identified research objectives (**ROs**) were the following: **RO1**, identify and analyse the existing literature's key research themes on the relationship between digitalization and organizational agility; **RO2**, detect this literature's main research gaps and limitations; **RO3**; identify the field's most promising future research avenues; **RO4**, identify and analyse the bidirectional nature of the relationship between digitalization and organizational agility.

The inclusion criteria (IC) and exclusion criteria (EC) used to select the papers for analysis appear in the first part of Figure 2. We selected the keywords (see IC1) on the basis of the main definitions of agility (Overby et al., 2006; Sherehiy et al., 2007; Walter, 2020) and digitalization (Legner et al., 2017; Sestino et al., 2020), as well as on a preliminary search in Google Scholar for the terms "agility" and "digitalization".

| | **Planning the review** | |
|---|---|---|
| 1. | Specification of research objectives | RO1. Identify and analyse the key research themes of the existing literature on the relationship between digitalization and organizational agility<br>RO2. Detect the main research gaps and limitations of this literature<br>RO3. Identify the most promising future research avenues in the field<br>RO4. Identify and analyse the bidirectional nature of the relationship between digitalization and organizational agility |
| 2. | Specification of inclusion criteria | IC1. Presence in abstract, keywords or title of the term "agil*" together with at least one of the following terms: "digitali?at*", "digiti?at*", "big data", "BD", "internet of thing*", "IoT", "digital transfor*", "digital technol*", "information technolog*", "ICT*", "Information Communication Technolog*", "Information and Communication Technolog*", "IT".<br>IC2. Studies published by 30 June 2021.<br>IC3. Studies published in English.<br>IC4. Studies limited to document type of journal article or review.<br>IC5. Studies already published or in press. |
| 3. | Specification of exclusion criteria | EC1. Studies that did not use "IT" as meaning "information technology".<br>EC2. Studies that did not focus on digitalization.<br>EC3. Studies that did not focus on organizational agility.<br>EC4. Studies that did not adopt a management perspective.<br>EC5. Studies that did not offer substantial insights into the relationship between agility and digitalization.<br>EC6. Studies that duplicate earlier articles. |
| 4. | Specification of database | Scopus. Cross-validation on Web of Science. |
| | **Undertaking of the review** | |
| 1. | Identification of search syntax | (TITLE-ABS-KEY(("digitali?at*"OR"digiti?at*"OR"big data"OR"BD"OR"internet of thing*"OR"IoT"OR"digital transfor*"OR"digital technol*"OR"information technolog*"OR"ICT*"OR"Information Communication Technolog*"OR"Information and Communication Technolog*"OR"IT")AND("agil*")))AND(LIMIT-TO(DOCTYPE,"ar")ORLIMIT-TO(DOCTYPE,"re"))AND(LIMIT-TO(LANGUAGE,"English")) |
| 2. | Execution of search query | 2,630 articles in the initial dataset as a result of the search query |
| 3. | Implementation of exclusion criteria | 1,762 articles removed due to EC1<br>230 articles removed due to EC2<br>197 articles removed due to EC3<br>142 articles removed due to EC4<br>93 articles removed due to EC5<br>12 articles removed due to duplication (EC6) |
| 4. | Undertaking citation chaining | 3 new articles included |
| 5. | Final dataset | 197 articles |
| | **Developing and reporting the review** | |
| 1. | Developing a concise research profile (see section 4) | Paper distribution per year/country/journal/author<br>Total number of citations per country/journal/author<br>Normalized total citations per country/journal/author<br>Average annual growth in the number of papers<br>Average number of citations per paper<br>Similarity analysis performed by using VOSviewer 1.6.10 software: 171 papers were found to be connected, which lead to a structure comprising three clusters |
| 2. | Identifying and analysing focal research themes (see section 5) | Green cluster: Information technology capabilities and organizational agility<br>Red cluster: Digitalization and supply chain agility<br>Blue cluster: Big data analytics capabilities, agility and performance |
| 3. | Identifying research gaps and exploring future research directions (see section 6) | Most relevant research gaps explored and discussed<br>Thirteen original research propositions developed and discussed |

Note for IC1: Asterisk ("*") allows for broadening the search by finding all words that start with the same letters. Question mark ("?") replaces a letter and allows for broadening the search by finding all words that contain any possible character in the place of the letter replaced by the question mark

**Figure 2.** Systematic literature review protocol and process

In line with the literature (Kraus et al., 2020), we chose the terms relating to the digitalization topic on the basis of the most relevant contributions in the field (Kraus et al., 2020). Given that the IoT and BDA are key enablers of business processes aimed at operationally exploiting digital technologies' potential (Pflaum and Gölzer, 2018; Sestino et al., 2020), we included the terms "big data" and "Internet of Things". In line with the literature (Delgado García et al., 2015; Kraus et al., 2020), we only selected journal articles or reviews (IC4) in English (IC3), already published or in press (IC5), and released by 30 June 2021 (IC2). The "*" and "?" operators were used as jolly characters to include as many lexical variants as possible.

A paper was excluded from our review if: (a) it contained "IT" in its title, abstract or keywords meaning the third-person singular pronoun and not information technology, (b) it did not focus on digitalization, (c) it did not focus on organizational agility, (d) it did not adopt a management point of view, (e) although it analysed both digitalization and agility, it did not offer substantial insights into the relationship between these two constructs, and (f) it was a duplicate of an earlier research result. The Scopus database, which the literature considers one of the ideal scientific databases for systematic literature reviews (Falagas et al., 2008; Kraus et al., 2020), was used for our search. We also implemented the search in another scientific database, Web of Science, for cross-validation purposes, without finding any new relevant documents.

On the basis of the selected keywords and the other four IC shown in Figure 1, we developed the following search syntax (step 1 of the execution stage):

*(TITLE-ABS-KEY(("digitali?at*"OR"digiti?at*"OR"big data"OR"BD"OR"internet of thing*"OR"IoT"OR"digital transfor*"OR"digital technol*"OR"information technolog*"OR"ICT*"OR"Information Communication Technolog*"OR"Information and Communication Technolog*"OR"IT")AND("agil*")))AND(LIMIT-TO(DOCTYPE,"ar")ORLIMIT-TO(DOCTYPE,"re"))AND(LIMIT-TO(LANGUAGE,"English"))*

Executing our search query allowed us to select an initial dataset of 2,630 papers.

In the following step, always in line with the literature ( e.g., Behera et al., 2019), three of the four authors independently undertook an autonomous review of each of the 2,630 documents (the titles, abstracts and keywords were reviewed in all cases, the full texts were reviewed when deemed necessary), in order to analyse their relevance by applying the six exclusion criteria specified in the planning stage (Tranfield et al., 2003). Krippendorf's alpha coefficient was calculated to measure the level of alignment between each author's obtained results. This was greater than 0.80, therefore supporting our selection protocol's robustness. A large number of papers (1,762) were excluded for having "IT" in their title, abstract or keywords, meaning the third-person singular pronoun and not information technology. Some papers were excluded for not delving into digitalization (230) or organizational agility (197), others for not adopting a management point of view (142), and still others (93) for not offering any substantial insights into the relationship between digitalization and agility, although they did analyse both these constructs. Twelve papers were excluded for being duplicates of earlier research results. This selection process reduced our dataset to 194 papers. Subsequently, in order to include additional relevant studies that the search syntax had not picked up, we applied the backward and forward citation chaining method (Webster and Watson, 2002). This led to the inclusion of another three new papers, bringing our dataset to 197 documents that provide an in-depth analysis of the relationship between digitalization and organizational agility from a management perspective. The fourth author reviewed these papers, along with the motivations for their inclusion, in order to crosscheck the selected dataset's robustness and undistorted nature (Behera et al., 2019).

In the third stage (developing and reporting the review), a concise bibliometric research profile was first developed in order to understand the structure and evolution of the literature that we have analysed. Bibliometric methods allow the effective interpretation of huge volumes of bibliographic information and the detection of research streams in specific fields (van Eck and Waltman, 2010), while simultaneously avoiding the potential bias that researchers' subjective interpretations usually generate (Zupic and Čater, 2015).

Further, in line with the existing literature (e.g., Bhatt et al., 2020; Khanra et al., 2021a, 2021b; Todeschini and Baccini, 2016), we calculated and interpreted certain relevant bibliometric indicators in order to analyse: the distribution of the selected papers per journal, author, country and year; the total number of citations; the normalized total number of citations per journal, author and country; the average annual growth in the number of papers; and the average number of citations per paper. Subsequently, we undertook a similarity analysis by using the bibliographic coupling algorithm (Van Eck et al., 2006; van Eck and Waltman, 2010), on the basis of which two contributions are considered coupled if they have one or more common third studies in their bibliography. This algorithm has proven useful and reliable regarding supporting the mapping of research fields and the recognition of research streams or trends (Boyack and Klavans, 2010). Among the different software packages available for implementing bibliometric analysis, namely BibExcel, CiteSpace, Gephi and VOSviewer (Bhatt et al., 2020), we used VOSviewer 1.6.10 software, which allowed us to build a graphical map in which each sphere represented a paper, and the papers are split into clusters as a function of the similarity of their references (Van Eck et al., 2006; van Eck and Waltman, 2010). The resulting cluster structure is a powerful instrument for interpreting the literature in a scientific dataset and identifying the research streams characterizing that literature (van Eck and Waltman, 2010). As a result of our similarity analysis, we found 171 connected papers in terms of their common references, forming a graphical structure comprising three clusters whose configurations seem quite well defined (see section 4).

Finally, three of the authors independently undertook three additional review processes of the full text of each of the 171 documents. The first process was aimed at identifying a series of focal research themes that the existing studies within each cluster addressed and assigning each paper to a specific topic. Each author performed an open, axial, and selective coding (Glaser et al., 1968) that: (i) abstracted the reviewed documents in order to develop categories, (ii) identified similarities and relationships between these categories, and (iii) selected the final research topics discussed in section 5. The second process was aimed at identifying the principal research gaps and future research directions. The authors also held a series of meetings to discuss on the research gaps and future research directions that each of them had previously independently identified. The final results of this second review process are presented and discussed in section 6. The third and last review process was aimed at selecting the most significant papers. We assigned a score to all 171 papers based on their significance and relevance for the topics to which they had been assigned. We held a series of meetings in order to reach agreement on the score that we would assign to each paper. This final process led us to select 70 representative papers (41% of the total dataset), on which we would subsequently focus (Gaur and Kumar, 2018; Tranfield et al., 2003) to discuss the focal research themes, to investigate the most relevant theoretical connections between these themes, to discover the most significant understudied topics and to propose a future research agenda. Again, we calculated, Krippendorf's alpha coefficient for each of these three final review processes to measure the agreement between the results that each author obtained. It was always greater than 0.80, thus confirming our review protocols' robustness.

## 4. Research profile and results of the VOS analysis

In the following figures and tables, we present some bibliometric indicators' dynamics that we consider useful for analysing and interpreting the structure and evolution of the literature that we have analysed.

Please note, that we use the following keys in this section:

1) NPs, an acronym for the Numb of Papers;

2) TCs, an acronym for the Total number of collected Citations;

3) Normalized Total Citations (NTCs), which represent a paper's total number of citations divided by the average number of total citations of all articles published in the same year and included in the dataset. This normalization corrects for the older documents having had more time to receive citations than more recent ones. The size of the bubbles in Figure 4 reflects each paper's NTC value;

4) AGR-NPs, an acronym for the Average annual Growth Rate of the NPs;

5) TCs/NPs an acronym for the average number of citations per paper.

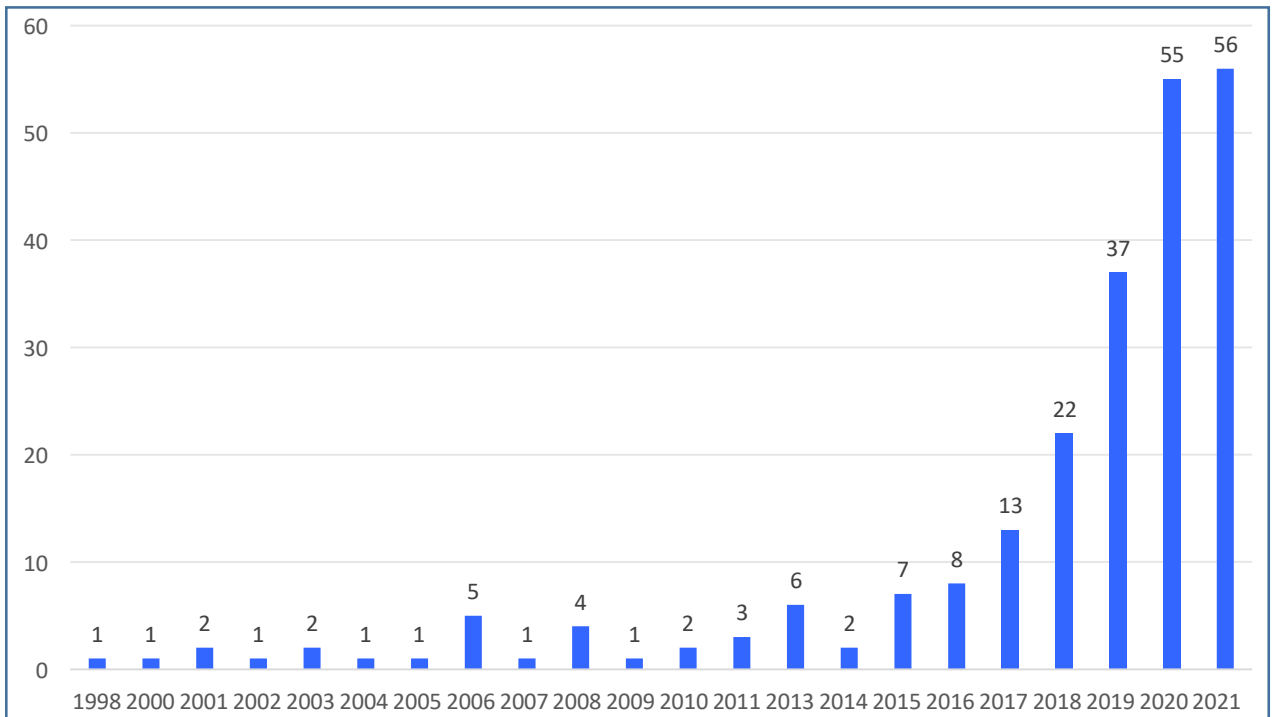

**Figure 3.** NPs per year

As shown, the number of publications per year has grown exponentially since 2017 (Figure 2), demonstrating that the scientific interest in the topic has recently increased strongly. This increase is also linked to the growing attention that the digital revolution and the leading technologies, on which the digital revolution is based, have attracted (Karimi and Walter, 2015; Vial, 2019).

Tables 1 and 2 show that the number and variety of journals that have published at least three studies, are relatively high. The main and specific scientific fields that the most relevant journals are: production, information management, supply chain management and technology management. The International Journal of Production Research published the largest number of studies (8). Among the most prolific journals, the presence of technology management (Technological Forecasting and Social Change), marketing (Industrial Marketing Management), and general management journals (e.g., the Management Research Review and Journal of Business Research). MIS Quarterly has the largest number of citations (almost 3,000), while the Journal of Business Research has the highest NTC value (17.0).

**Table 1. Papers and citations per journal (top 10 journals)**

| Journals ranked by NPs (top 10 journals) | | | |
|---|---|---|---|
| Internat. Journal of Production Research | 8 | Industrial Management and Data Systems | 4 |
| Journal of Business Research | 7 | Technolog. Forecasting and Social Change | 4 |
| Internat. Journal of Information Management | 7 | Decision Support Systems | 3 |
| Internat. Journal of Supply Chain Management | 6 | MIS Quarterly | 3 |
| Management Research Review | 4 | Internat. Journal of Production Economics | 3 |
| **Journals ranked by TCs (top 10 journals)** | | | |
| MIS Quarterly | 2,940 | European Journal of Operational Research | 416 |
| Internat. Journal of Production Economics | 1,009 | Decision Support Systems | 413 |
| Internat. Journal of Production Research | 904 | Internat. Journal of Information Manag. | 406 |
| European Journal of Information Systems | 700 | Information Systems Research | 359 |
| Journal of Business Research | 431 | Journal of Strategic Information Systems | 328 |
| **Journals ranked by NTCs (top 10 journals)** | | | |
| Journal of Business Research | 17.0 | Long Range Planning | 7.4 |
| Internat. Journal of Production Economics | 16.4 | Journal of Internat. Management | 7.2 |
| Internat. Journal of Information Management | 13.9 | Journal of Strategic Information Systems | 6.9 |
| Internat. Journal of Production Research | 11.7 | Information Systems Research | 6.7 |
| Industrial Marketing Management | 7.7 | Technolog. Forecasting and Social Change | 5.7 |

**Table 2. Papers and citations per authors (top 10 authors)**

| Authors ranked by NPs (top 10 authors) | | | |
|---|---|---|---|
| Gunasekaran, A. | 8 | Adeleye, E.O. | 3 |
| Sambamurthy, V. | 5 | Mandal, S. | 3 |
| Panda, S. | 5 | Yusuf, Y.Y. | 3 |
| Rath, S.K. | 5 | Zhang, J. | 3 |
| Papadopoulos, T. | 4 | Dubey R. | 3 |
| **Authors ranked by TCs (top 10 authors)** | | | |
| Sambamurthy, V. | 2,870 | Dubey R. | 137 |
| Gunasekaran, A. | 1,644 | Zhang, J. | 62 |
| Papadopoulos, T. | 741 | Panda, S. | 44 |
| Adeleye, E.O. | 449 | Rath, S.K. | 44 |
| Yusuf, Y.Y. | 449 | Mandal, S. | 40 |
| **Authors ranked by NTCs (top 10 authors)** | | | |
| Gunasekaran, A. | 25.7 | Yusuf, Y.Y. | 5.8 |
| Dubey R. | 13.9 | Zhang, J. | 3.9 |
| Sambamurthy, V. | 12.1 | Mandal, S. | 2.1 |
| Papadopoulos, T. | 12.1 | Panda, S. | 1.0 |
| Adeleye, E.O. | 5.8 | Rath, S.K. | 0.9 |

The distribution of studies per author (Table 2) reveals a significant number of very productive scientists (ten authors published at least three pieces, with five of them producing four papers or more). Sambamurthy is the most prolific author with the largest number of collected citations (almost 2,900) by far, while Gunasekaran has the highest NTC value (25.7).

**Table 3.** Papers and citations per country (top 10 countries)

| Countries ranked by NPs (top 10 countries) | | | |
|---|---|---|---|
| United States | 49 | France | 9 |
| United Kingdom | 31 | Indonesia | 8 |
| India | 28 | Italy | 10 |
| China | 23 | Australia | 7 |
| Finland | 10 | Hong Kong | 6 |
| **Countries ranked by TCs (top 10 countries)** | | | |
| United States | 7,455 | France | 264 |
| United Kingdom | 2,192 | Italy | 253 |
| Hong Kong | 1,255 | Australia | 143 |
| China | 781 | Finland | 116 |
| India | 331 | Indonesia | 27 |
| **Countries ranked by NTCs (top 10 countries)** | | | |
| United States | 83.3 | Finland | 17.4 |
| United Kingdom | 54.2 | Italy | 16.7 |
| India | 22.9 | Hong Kong | 14.9 |
| China | 22.2 | Australia | 11.6 |
| France | 20.1 | Indonesia | 2.9 |

Table 3 highlights that 40% of the scientific production comes from authors in the United States (49) or the United Kingdom (31), although Chinese and Indian authors are also quite prolific (23 and 28 papers, respectively). The United States has the largest number of collected citations (more than 7,000) and the highest NTC value (83.3) by far.

**Table 4.** Top 20 papers (ranked on the basis of TCs)

| Authors | Title | Year | TCs | NTCs |
|---|---|---|---|---|
| Sambamurthy V., Bharadwaj A., Grover V. | Shaping agility through digital options: Reconceptualizing the role of information technology in contemporary firms | 2003 | 1,855 | 1.8 |
| Overby E., Bharadwaj A., Sambamurthy V. | Enterprise agility and the enabling role of information technology | 2006 | 565 | 3.0 |
| Tallon P.P., Pinsonneault A. | Competing perspectives on the link between strategic information technology alignment and organizational agility: Insights from a mediation model | 2011 | 562 | 1.3 |
| Wang G., Gunasekaran A., Ngai E.W.T., Papadopoulos T. | Big data analytics in logistics and supply chain management: Certain investigations for research and applications | 2016 | 523 | 5.1 |
| Lu Y., Ramamurthy K. | Understanding the link between information technology capability and organizational agility: An empirical examination | 2011 | 523 | 1.2 |
| Gunasekaran A. | Agile manufacturing: Enablers and an implementation framework | 1998 | 437 | 1.0 |
| Swafford P.M., Ghosh S., Murthy N. | Achieving supply chain agility through IT integration and flexibility | 2008 | 434 | 3.7 |
| Liu H., Ke W., Wei K.K., Hua Z. | The impact of IT capabilities on firm performance: The mediating roles of absorptive capacity and supply chain agility | 2013 | 349 | 3.1 |
| Yusuf Y.Y., Gunasekaran A., Adeleye E.O., Sivayoganathan K. | Agile supply chain capabilities: Determinants of competitive objectives | 2004 | 339 | 1.0 |
| Arzu Akyuz G., Erman Erkan T. | Supply chain performance measurement: A literature review | 2010 | 222 | 1.4 |

| Chakravarty A., Grewal R., Sambamurthy V. | Information technology competencies, organizational agility, and firm performance: Enabling and facilitating roles | 2013 | 208 | 1.8 |
|---|---|---|---|---|
| Mikalef P., Pateli A. | Information technology-enabled dynamic capabilities and their indirect effect on competitive performance: Findings from PLS-SEM and fsQCA | 2017 | 204 | 4.2 |
| Ngai E.W.T., Chau D.C.K., Chan T.L.A. | Information technology, operational, and management competencies for supply chain agility: Findings from case studies | 2011 | 179 | 0.4 |
| Larson D., Chang V. | A review and future direction of agile, business intelligence, analytics and data science | 2016 | 166 | 1.6 |
| Swafford P.M., Ghosh S., Murthy N.N. | A framework for assessing value chain agility | 2006 | 161 | 0.9 |
| Lee O.-K., Sambamurthy V., Lim K.H., Wei K.K. | How does IT ambidexterity impact organizational agility? | 2015 | 151 | 4.8 |
| Côrte-Real N., Oliveira T., Ruivo P. | Assessing business value of Big Data Analytics in European firms | 2017 | 146 | 3.0 |
| Breu K., Hemingway C.J., Strathern M., Bridger D. | Workforce agility: The new employee strategy for the knowledge economy | 2002 | 145 | 1.0 |
| Warner K.S.R., Wäger M. | Building dynamic capabilities for digital transformation: An ongoing process of strategic renewal | 2019 | 138 | 7.4 |
| Paulraj A., Chen I.J. | Strategic buyer-supplier relationships, information technology and external logistics integration | 2007 | 130 | 1.0 |

Table 4 shows the top 20 papers in term of total citations (TCs). It also presents, in the last column, the values of NTCs reported by each paper. While the study by Sambamurthy et al. (2003) is by far the most cited (with more than 1,850 TCs), the one by Warner and Wäger (2019) is the one which received the largest number of citations in relation to the time elapsed from its publication (NTCs equal to 7.4).

Trough bibliographic coupling two articles citing a publication are coupled because high instances of mutual reference suggest an intellectual capital common to both (Khanra et al., 2021a, 2021b).
The clustering structure resulting from a similarity analysis based on the bibliographic coupling algorithm and implemented by means of VOSviewer 1.6.10 software (Figure 4) reveals the presence of three thematic clusters whose configurations seem rather well outlined and significant inter-cluster connections.
The papers in the blue cluster view big-data analytics capabilities (BDACs) and other digital technologies, such as IoT, artificial intelligence (AI), and social media platforms, as crucial drivers of company agility and, through the latter, of company performance. The red cluster aggregates studies focused on the relationship between digital technologies and agility at the supply chain (SC) level. Finally, the green cluster assembles studies focused on the role of information technology capabilities (ITCs) in enhancing organizational agility (OA) and on the ITCs-OA relationship's impact on firm competitiveness and innovation performance.

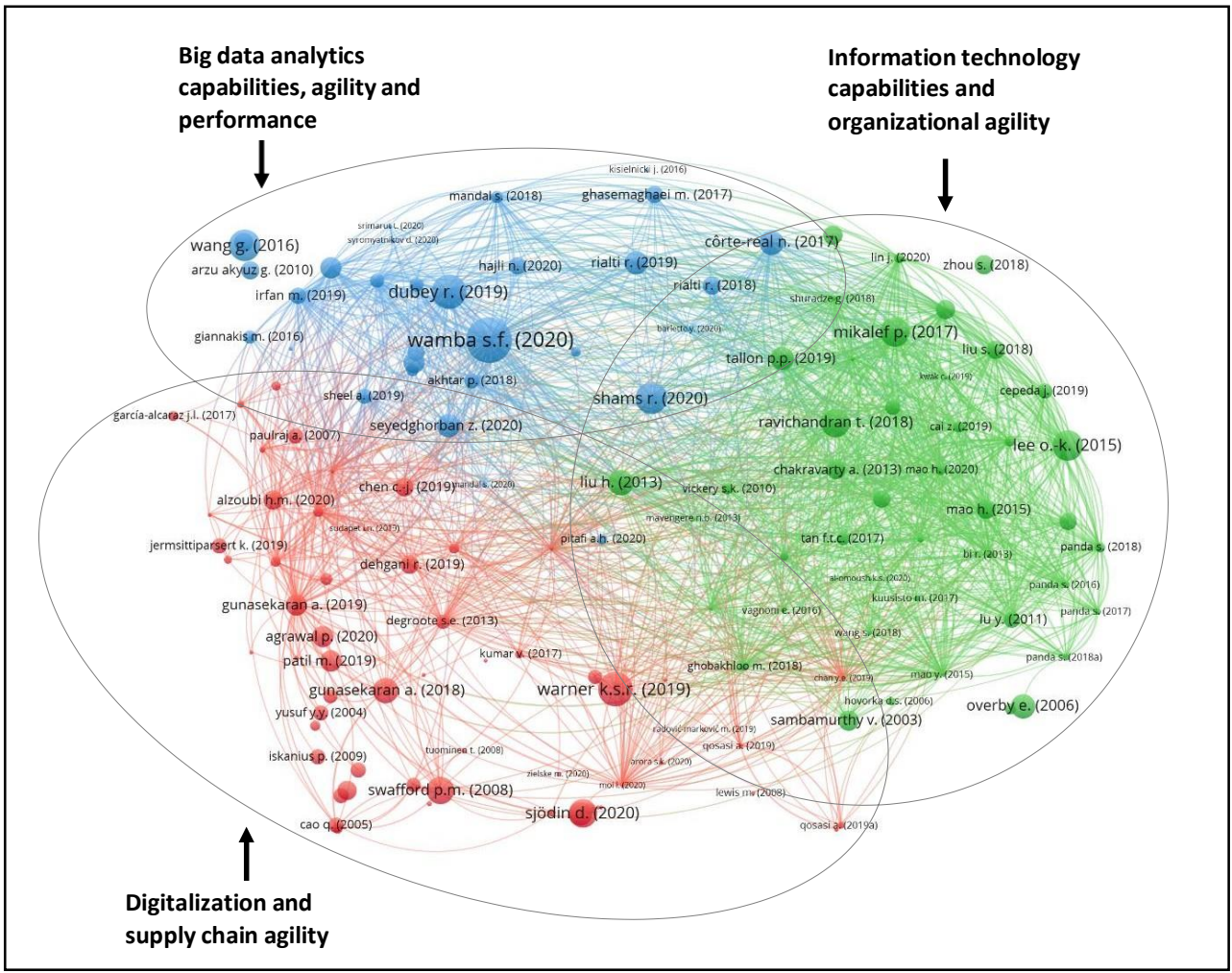

**Figure 4**. The clustering structure emerged from the VOS analysis

In Figure 5 and Table 5 we present the development over time and a few bibliometric indicators characterizing each cluster.

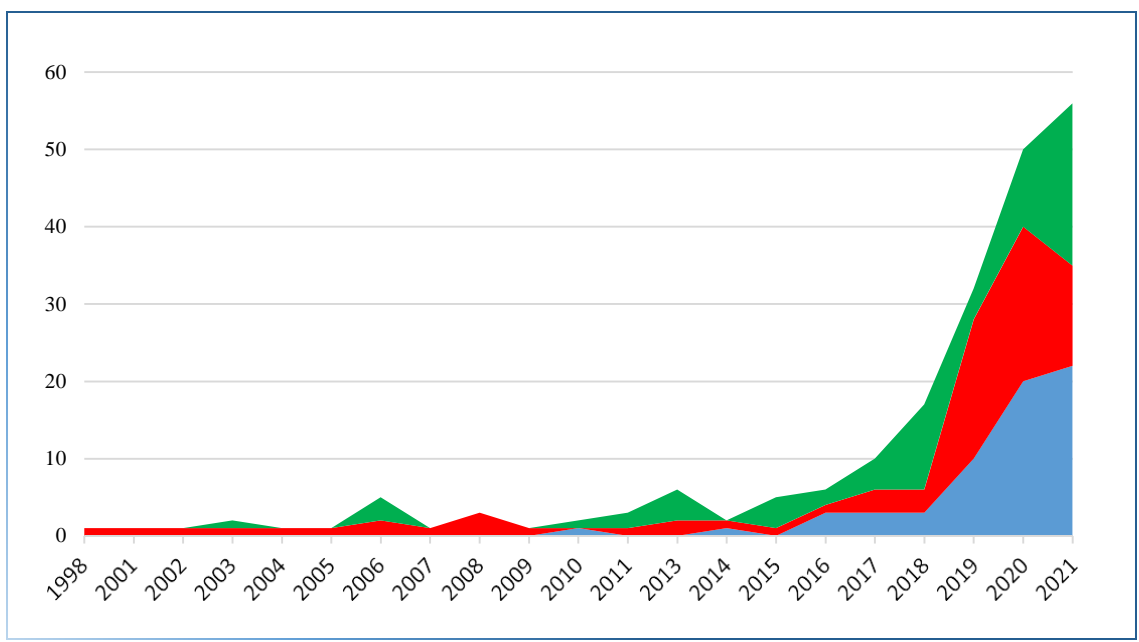

**Figure 5.** Papers per year-cluster

All the clusters have shown a clear and strong growth trend during the last three years, with the green starting to grow in 2018 and the other two since 2019.

Table 5. Bibliometric indicators characterizing each cluster

|  | Red Cluster | Blue Cluster | Green Cluster | Totals |
|---|---|---|---|---|
| **Number of Papers (NPs)** | 68 | 48 | 55 | 171 |
| **Total Citations (TCs)** | 3,024 | 1,753 | 5,210 | 9,987 |
| **Normalized Total Citations (NTCs)** | 66.9 | 72.7 | 52.1 | 191.7 |
| **Number of Papers' Average Growth Rate (AGR-NPs)** | +72.7% | +49.2% | +57.6% | +61.0% |
| **Total Citations/Number of Papers (TCs/NPs)** | 44.5 | 36.6 | 94.7 | 58.4 |

The green cluster has collected the largest number of TCs by far (70% more than the red cluster and almost double that of the blue one), but is last in terms of NTCs (52.1, -28% compared to the blue cluster and -22% compared to the red cluster). With exception of the last three years of our analysis (2019, 2020, and 2021), this cluster's NPs was three times more than those of the red cluster and +30% of the blue one. These findings suggest that the studies focused on information technology capabilities' role in improving OA (which constitutes this cluster's thematic focus) are the more developed research stream within the literature that we have reviewed.

The red cluster occupies a middle position in terms of both its TCs (3,024) and its NTCs (66.9). It is the first cluster in terms of both NPs (68) and its average annual growth rate is +72.7% compared to the entire dataset's value of +61.0%), which shows that the digitalization-agility interconnections' analysis at the supply chain level takes a central position within the literature object. The blue cluster has the lowest average number of citations per paper (36.6) and the smallest number of NPs (48) and TCs (1,753). Nevertheless, this cluster is simultaneously the first in terms of its NTCs (72.7), which demonstrates that the interest in the relationship between big-data analytic capabilities and OA (which is this cluster's core topic) is increasing at a higher than average rate.

## 5. Results of the systematic review

We present the results of our literature review in the following sections. Table 6 summarizes the main thematic areas that were the object of analysis within each cluster.

Table 6. Main topics per cluster

| Main Topics | References |
|---|---|
| **GREEN CLUSTER: INFORMATION TECHNOLOGY CAPABILITIES AND ORGANIZATIONAL AGILITY** | |
| Information technology capabilities as enablers of organizational agility | Gao et al., (2020); Liu et al., (2018); Lu and K. (Ram) Ramamurthy, (2011); Tallon and Pinsonneault (2011); Zhen et al. (2021). |
| The enhancing factors of the relationship between information technology capabilities and organizational agility | Lin et al., (2020); Mao et al., (2015); Mao et al., (2021); Panda and Rath, (2016, 2017); Baloch et al. (2018); Panda (2021). |
| Combining information technology capabilities and organizational agility to gain competitive advantages and enhance firm performance | Liu et al., (2013); Mikalef and Pateli, (2017); Vagnoni and Khoddam (2016); Martínez-Caro et al., (2020); Sambamurthy et al., (2003). |
| The mediating role of innovation capabilities and culture in the relationship between information technology capabilities and organizational agility | Cai et al., (2019); Cepeda and Arias-Pérez (2019); Ravichandran (2018); Nwankpa and Merhout (2020). |
| **RED CLUSTER: DIGITALIZATION AND SUPPLY CHAIN AGILITY** | |
| Digitalization and supply chain agility | Agrawal et al., 2019; Bargshady et al. (2016); Brenner, (2018); Chan et al. (2019); Mak and Shen, (2020) Shiranifar et al. (2019); Warner and Wager, (2019); Choudhury et al., (2021). |
| Digital-based supply chain agility and firm performance | Alzoubi and Yanamandra, 2020, Chen, 2019; Degroote and Marx (2013); Garcia-Alcaraz et al. (2017); Swafford et al., (2008). |
| Digital organizational culture and supply chain agility | Dehgani and Jafari Navimipour, (2019); Ngai et al., (2011); Malekifar et al., (2014); Jermsittipar and Wajeetongratana (2019). |
| Digital-based supply chain agility: manufacturing and logistic perspectives | Gunasekaran et al., (2019); Gunasekaran et al., (2018); Zielske and Held, (2020); Paulraj and Chen, (2007). |
| **BLUE CLUSTER: BIG DATA ANALYTICS CAPABILITIES, AGILITY AND PERFORMANCE** | |
| Big data analytics capabilities, ambidexterity and agility | Fosso-Wamba et al., (2020); Rialti et al., (2018); Rialti et al., (2019); Shams et al., (2021). |
| Big data analytics capabilities, agility and performance | Asrini et al., (2020); Côrte-Real et al., (2017); Hajli et al., (2020); Stylos et al., (2021); Von Alberti-Alhtaybat et al., (2019). |
| The role of organizational competences and learning culture | Barlette & Baillette, (2020); Fachnurrisa et al., (2020); Ghasemaghaei et al., (2017); Kisielnicki & Misiak, (2016); Kozarkiewicz (2020). |
| Big data analytics capabilities and supply chain agility | Christopher & Ryals, (2014); Dubey et al., (2019); Giannakis & Louis, (2016); Mandal, (2018); Meriton et al., (2020); Raut et al., (2021); Wang et al., (2016). |
| The new digital solutions as drivers of organizational agility | Ahn, (2020); Akhtar et al., (2018); Calatayud et al., (2019); Nandi et al., (2020); Pitafi et al., (2020); Rane & Narvel, (2019); Sjödin et al., (2021); Sheel & Nath, (2019). |

*5.1 Green cluster: information technology capabilities and organizational agility*

Green cluster aggregates studies focusing on the role of ITCs in improving OA and on the impact of the ITCs-OA relationship on firm competitiveness and innovation performance.

*5.1.1 Information technology capabilities as enablers of organizational agility*

Superior ITCs, defined as the company's ability to leverage IT to develop business strategies and processes (Lu and K. (Ram) Ramamurthy, 2011), represents a significant enabler of OA (Gao et al., 2020; Liu et al., 2013; Mao et al., 2015; Martínez-Caro et al., 2020; Overby et al., 2006). Technical ITCs are constituted by the combination of IT flexibility, i.e., the company's ability to evolve IT infrastructures, and IT integration, which is the extent to which IT and information is diffused among company departments (Saraf et al., 2007). Since both IT flexibility and IT integration have a crucial role in improving OA (Liu et al., 2018), firms should be able to develop modular IT components and integrate such components into the entire organizational environment (Gao et al., 2020). Liu et al. (2018) demonstrate that a flexible and integrated cloud-based IT infrastructure allows organizations to effectively deploy IT applications, respond to customers' demands and connect with new partners. However, technical ITCs can be effectively leveraged to enable OA, but only if a company has sufficient managerial ITCs, which are defined as the abilities to align an IT infrastructure and a business strategy, and use the first to support its business goals and create new opportunities (Gao et al., 2020; Lu and K. (Ram) Ramamurthy, 2011; Panda and Rath, 2017, 2016; Tallon and Pinsonneault, 2011). Gao et al. (2020) suggests that in the presence of an adequate IT business spanning capability (i.e., the capability to integrate business and IT strategic planning), IT flexibility has a stronger effect on OA than IT integration. In other words, only organizations that can continuously adapt their IT components to emerging changes can fully leverage their managerial ability's OA effect in order to boost the information sharing between the IT and business managers and employees. Panda and Rath (2017) find that organizations' human ITCs have a positive impact on their sensing and responding agility. Lu and K. (Ram) Ramamurthy (2011) conceptualize ITCs as a construct based on three dimensions: IT infrastructure capabilities, IT business spanning capabilities, and IT proactive stance. These dimensions represent a company's ability to continuously discover new ways of developing IT innovations or leveraging existing IT resources to create value. IT ambidexterity (IT exploration and IT exploitation) has recently been proved to be positively related to OA. Specifically, by leveraging organizational inertia theory and literature towards IT-enabled agility, Zhen et al. (2021) find that both IT exploration (a firm's ability to develop new IT capabilities) and, to a greater extent, IT exploitation (a firm's ability to effectively leverage existing IT technologies and resources) have a positive effect on OA, and that both partially mediate and weaken organizational inertia's negative effect on OA.

*5.1.2. The factors enhancing the relationship between information technology capabilities and organizational agility*

Other studies investigate the mediating or moderating roles that information intensity and environmental uncertainty (Mao et al., 2015), environmental dynamism (Baloch et al., 2018; Chakravarty et al., 2013; Lin et al., 2020), IT spending (Panda and Rath, 2017, 2016), and organizational variables, such as absorptive capacity (Mao et al., 2021) and operational dynamic capabilities (Baloch et al., 2018), play in the ITCs-OA relationship. In turbulent environments firms have to process huge volumes of data to be sufficiently agile, as well as to sense environmental changes and respond to them effectively. Consequently, environmental uncertainty dynamism and complexity strengthen ITCs' effect on OA (Baloch et al., 2018; Mao et al., 2015; Panda, 2021), while environmental hostility weakens it (Panda, 2021). Furthermore, high levels of product or service information intensity generate the need to collect, elaborate, and turn a large amount of data and information into knowledge. Consequently, the higher the level of information intensity, the greater IT's impact on OA (Mao et al., 2021; Mao et al., 2015). Panda and Rath (2017, 2016) maintain that firms should spend an adequate amount of money on IT assets in order to generate the ITCs required to reach high levels of agility. However, if firms invest mainly in IT infrastructure rather than IT managerial assets, IT spending does not have a significant effect on their overall agility. Baloch et al. (2018) find that operational dynamic capabilities mediate ITCs' influence on OA significantly and positively, while Mao et al. (2021) show that absorptive capacity acts as a mediator in the relationship between IT knowledge and both market capitalizing agility and operational adjustment agility.

*5.1.3. Combining information technology capabilities and organizational agility to gain competitive advantages and enhance firm performance*

While the above-mentioned studies aim to explore and measure the IT-related factors that directly or indirectly impact OA, another group of contributions focuses on OA and IT as antecedents of firm performance and competitive advantages. Mikalef and Pateli (2017) find that OA mediates the IT dynamic capabilities' effects on firm performance. Specifically, IT dynamic capabilities enhance market capitalizing agility (therefore providing first-mover advantages regarding product customization) and operational adjustment agility (allowing the production pace to quickly adapt to the customer requirements). In turn, this enhancement leads to higher customer retention and superior competitive performance. Martìnez-Caro et al. (2020) demonstrate that IT assimilation capacity, defined as the extent to which technology use diffused across a company has become routinized, has a positive and direct impact on OA and a positive indirect effect on the firm performance. Liu et al. (2013) find that OA mediates IT assimilation capabilities' effects on the firm performance, while Vagnoni and Khoddami (2016) demonstrate that IT dynamics capabilities improve firms' strategic agility, which, in turn, has a positive effect on the company competitiveness.

*5.1.4. The mediating role of innovation capabilities and culture in the relationship between information technology capabilities and organizational agility*

Ravichandran (2018) finds that, while ITCs allow the development of new business models, only an organizational culture that strongly stimulates innovation and risk taking will enable the effective exploitation of the IT competences required to obtain a high level of OA and a strong ability to detect and exploit market opportunities. Cai et al. (2019) define an "innovative climate" as an environmental factor that encourages employees to generate new ideas, thoughts and changes; these authors find that such a climate enhances market agility, which they regard as the ability to continuously develop new products. Cepeda and Arias-Pérez, (2019) explore open innovation's mediating role in the ITCs-OA relationship directly, finding that open innovation capabilities allow a company to integrate inbound and outbound knowledge flows effectively, which subsequently accelerates innovation and allows a company to gain agility when responding to context changes. Finally, Nwankpa and Merhout, (2020) suggest that digital investment and OA are key enablers of IT innovation and that only firms endowed with superior organizational agility can effectively leverage digital technology investments, capture insights and opportunities from a business climate dominated by digital platforms, and fully convert digital investment's potential into IT innovation.

*5.2. Red cluster: digitalization and supply chain agility.*

The red cluster aggregates studies focused on the relationship between digital technologies and agility at the supply chain (SC) level.

*5.2.1. Digitalization and supply chain agility*

SC agility (SCA) is the SC's ability to sense and respond rapidly and effectively to environmental changes. In order to do so, SCs need to be market sensitive (i.e., effectively connected to customer trends), virtual (i.e., able to exploit digital technologies and platforms to share information between all the chain members in real time), network-based (i.e., based on the interconnected usage of members' strengths), and process-aligned (i.e., characterized by a high level of process synchronization between all the network players) (Bargshady et al., 2016). Agility is the core mechanism that facilitates SC digital transformations by forcing companies to innovate their business models, collaborative approaches, and culture (Warner and Wäger, 2019) and by allowing them to introduce a partnering, co-creating, and value-sharing logic (Brenner, 2018). At the same time, the development of an integrated and flexible digital IT infrastructure is crucial to fully leverage an agile SC's typical characteristics, such as its fast new product development, flexible planning, low lead times and costs, high product and service quality, and high levels of customer satisfaction. In this regard, digitalization offers new opportunities to develop the SC triple-A (agility, adaptability and alignment) approach further and to complete the transition from a production- to a fully demand-driven logic (Mak and Shen, 2021). Several studies (Bargshady et al., 2016) demonstrate that there is a positive and strong relationship between IT capabilities and SCA, and that a balanced and coherent IT and knowledge capabilities co-evolution influences the latter positively (Chan et al., 2019; Shiranifar et al., 2019). Choudhury et al. (2021) have recently demonstrated that an SC's efficiency and agility depend very much on the SC's ability to overcome traditional structures by designing and implementing major SC restructuring processes based on an effective exploitation of the emerging digital technologies. However, the process of SC digital transformation can only be effective if certain barriers are overcome, such as the lack of adequate digital skills and high implementation costs (Agrawal et al., 2019).

*5.2.2. Digital-based supply chain agility and firm performance*

SCA has positive effects on the performance of all companies that are part of the SC (Alzoubi and Yanamandra, 2020; Chen, 2019; DeGroote and Marx, 2013; Swafford et al., 2008). Alzoubi and Yanamandra (2020) find that digital technologies favour SCA, as they facilitate collaborative decision-making's development and the exchange and sharing of procedures, databases and applications, therefore leading to all SC actors experiencing greater levels of productivity, customer satisfaction and competitive performance. Similarly, DeGroote and Marx (2013) find that although digital technologies only have a slight effect on information timeliness, they impact information quality and accuracy meaningfully, allowing the implementation of effective and well-coordinated SC planning processes, which, in turn, have a positive impact on customer satisfaction and financial performance levels. The SC ability to leverage digital technologies effectively in order to respond promptly to market threats and opportunities (strategic flexibility) and to reconfigure manufacturing resources timely (manufacturing flexibility), leads to reduced SC lead times and to the effective adaptation of production volumes and variety to emerging market dynamics and customer expectations (Swafford et al., 2008). Trust between SC partners is an important amplifier of the digital-based supply chain agility's impact on firm performance, because it allows the SC members to achieve high levels of coordination and collaboration, which reduces the costs, improves the product innovation processes and creates more value for the customers (Chen, 2019). Finally, the widespread use of IT digital tools favours SC actors' ability to effectively access and share information, which are key to ensuring on-time deliveries, high levels of product customization, customer satisfaction, and short product development and production times. Managers should therefore encourage ICT implementation and ensure that their employees undergo adequate ICT training to maximize the overall SC value creation potential, and the chances that their firms will remain competitive over time (Garcia-Alcaraz et al., 2017).

*5.2.3. Digital organizational culture and supply chain agility*

To develop agility, SC organizations have to be endowed with adequate IT skills and knowledge. Dehgani and Jafari Navimipour (2019) demonstrated that to effectively leverage IT technologies for SCA, all SC actors should have adequate knowledge and skills concerning user need dynamics, on the one hand, and software development, project management, hardware and software compatibility issues, on the other hand. It is crucial that firms increasingly train their managerial and operational personnel on how to use and exploit all state-of-the-art IT digital tools. Ngai et al. (2011) argue that IT competences, which refer to IT integration and flexibility knowledge; operational competences, which concern SC integration and flexibility know-how; SC learning orientation; and management competencies, which refer to management's IT vision, are all necessary, interlinked and mutually supportive components of the firm's competence on which an effective SCA should be based. Organizational culture, which refers to an articulated combination of roles, values, assumptions, and symbols that characterize a company's vision of the world, moderates the relationship between IT competences and SCA positively (Malekifar et al., 2014). Only when the IT is sufficiently dynamic and strong, and the enterprise culture is agility oriented, is it possible to effectively leverage the IT competencies that managers and employees have in order to obtain high levels of SCA (Jermsittiparsert and Wajeetongratana, 2019).

*5.2.4. Digital-based supply chain agility: manufacturing and logistics perspectives*

Achieving a pronounced digital-based SC ability to quickly respond to market instability and the product complexity evolution requires close supervision and active coordination of the activities of all of the SC actors, such as the suppliers, manufacturers and logistic distributors. This is especially true in respect of manufacturing activities. In fact, digital technologies and big data analytics capabilities (Gunasekaran et al., 2018) are fundamental enablers of agile manufacturing, which simultaneously minimizes the time to market and the lead times, while maximizing production flexibility and the product customization level (Gunasekaran et al., 2019). Furthermore, consumers' increased expectation regarding the rapid delivery of goods and services has recently made logistics companies' role much more critical for the entire SC's overall agility and competitiveness. Paulraj and Chen (2007) found that logistics companies' digital-based agility depends strongly on the supplier-buyer relationships' quality and on the level of the IT chain integration. Zielske and Held (2020) demonstrate that logistics start-ups can apply digital-based agile practices effectively in competitive contexts characterized by high levels of market insecurity. In turn, these practices enhance the responsiveness to changing priorities and demands, accelerate product delivery and help realize high levels of coordination between IT and business departments.

*5.3. Blue cluster: big data analytics capabilities, agility and performance.*

IT infrastructure is a set of shared tangible IT resources enabling business applications. Of these resources, platform technologies, networks and telecommunications, critical data and data processing applications play a crucial role in firm competitiveness (Broadbent et al., 1999). The term big data (BD) refer to complex and large datasets that cannot be analysed by using standard statistical models (Ghasemaghaei et al., 2017; Mandal, 2018). The papers in the blue cluster assume that, in the digitalization era, BD and other emerging digital technologies linked to it (such as IoT, AI, blockchain and social media platforms) are primary sources of a company's competitive advantage and that big data analytics capabilities (BDACs) have a profound impact on the company agility and, therefore, on the company performance.

*5.3.1. Big data analytics capabilities, ambidexterity and agility*

A first group of contributions underscores that BDACs affect the company ambidexterity, which, in turn, contributes to firms' agility (Rialti et al., 2019, 2018). BDACs have a positive impact on ambidextrous capabilities by favouring IT infrastructure flexibility (Shams et al., 2021) and the company's ability to track all the relevant changes in the internal and external environment (Fosso Wamba et al., 2020), manage supply chain disruptions (Fosso Wamba et al., 2020), and rapidly and effectively respond to customers' needs (Rialti et al., 2019). However, using complex IT infrastructures for BD collecting could also be a limitation for the company agility when adopting new IT technologies or infrastructures, because the digital shift could cause data loss or unavailability (Shams et al., 2021).

*5.3.2. Big data analytics capabilities, agility and performance*

Several of this cluster's contributions focus on the relationship between BDACs, agility, company marketing, financial performance and competitive advantage. BDA tools support effective marketing decisions, such as those concerning new advertising campaigns, new product launches or market monitoring (Hajli et al., 2020). This is true in B2C industries, such as tourism, and in B2B contexts, such logistics, in which companies can use BDA effectively to gain insights into their customers' needs and behavioural patterns as well as to leverage these insights to effectively tailor their value proposal for clients (Stylos et al., 2021; von Alberti-Alhtaybat et al., 2019). Hajili et al. (2020) find that BDACs are important to achieve high market performance from new product launches, as they allow firms to effectively sense and respond to the continuous changes in customer needs. Adopting a knowledge-based view (KBV), Côrte-Real et al. (2017) find that BDA enhances firm agility and allows to achieve a superior financial performance by providing easy access to critical knowledge and information and enabling managers to make and change decisions rapidly and effectively, and (Asrini et al., 2020).

*5.3.3. The role of organizational competences and learning culture*

Exploiting BDA successfully to enhance company agility, also requires developing an organizational learning-oriented culture (Barlette and Baillette, 2020), which agile leadership can develop effectively (Fachnurrisa et al., 2020). Ghasemaghaei et al. (2017) find that for BDA to affect company agility positively, an adequate fit should first be achieved between the analysed data, the data analysis tools, their functionalities, the tasks to be accomplished, as well as the digital and agility-oriented competencies and learning culture. In particular, agility-oriented and digital competencies have been found essential for leveraging BDA to implement business intelligence projects (Kisielnicki and Misiak, 2016, Kozarkiewicz 2020).

*5.3.4. Big data analytics capabilities and supply chain agility*

Other contributions by this cluster focus on BDACs' role as a driver of SCA. Indeed, in order to rapidly and effectively address the SC logistics issues that an ever-changing and turbulent environment generates, requires access to the precious insights that BDA's appropriate usage generates (Giannakis and Louis, 2016; Mandal, 2018). While Dubey et al. (2019) identify organizational flexibility as a moderator of the relationship between BDACs and SCA, Raut et al. (2021) find that BDA mediates the impact of several organizational and management practices (such as lean management, environmental, and total quality management practices) on SCA positively. Wang et al. (2016) go beyond the impact of BDACs on SCA, finding that the latter largely contributes to a superior competitive advantage due to the high demand forecast precision and inventory management effectiveness that BDA's proper usage ensures, thus paving the way for new demand chain forecasting theories and best practices (Christopher & Ryals, 2014; Meriton et al., 2020).

*5.3.5. The new digital solutions as drivers of organizational agility*

In addition to BD, other digital assets linked to it could be valuable drivers of company agility, such as IoT (Ahn, 2020; Akhtar et al., 2018; Rane and Narvel, 2019), AI (Calatayud et al., 2019; Sjödin et al., 2021), blockchain (Nandi et al., 2020; Sheel and Nath, 2019) and even social media (Pitafi et al., 2020). IoT solutions make innovation processes agile by enhancing information sharing and communication (Ahn, 2020). Furthermore, IoT installations' usage for the real-time control of production lines increases production agility by allowing effective predictive maintenance and the quick identification and repair of mechanical failures, thus avoiding all manual checks and production stoppages that would otherwise be mandatory (Rane and Narvel, 2019). The usage of AI self-learning algorithms (Calatayud et al., 2019) and blockchain technologies (Rane and Narvel, 2019), which enable a greater degree of supply chain resilience (Nandi et al., 2020) as well as higher levels of communication effectiveness and trust between SC actors, enhance IoT solutions' contribution to production systems' agility (Sheel and Nath, 2019). Finally, companies currently widely and increasingly adopt tools such as DingTalk, Slack, Microsoft Teams, Trello, and Yammer to maximize their business processes' agility and to effectively support employee cooperation and knowledge sharing (Pitafi et al., 2020).

## 6. Discussion and research propositions

Table 7 synthesizes the most relevant understudied topics that emerged from our review. The references presented in the second column identify the papers that have directly or indirectly inspired the research gaps synthesized in the first column.

Some examples of addressable research gaps in the green cluster concern our understanding of the level of complementarity between the different digital capabilities required to optimize company agility in different environmental contexts (Gao et al., 2020; Liu et al., 2018), of how IT-based assimilation capabilities can improve organizational agility (Mao et al., 2021; Martìnez-Caro, 2020), and of which non-IT capabilities could be leveraged to optimize the IT-OA relationship and how this should be done (Ravichandran, 2018). Furthermore, the ever-increasing environmental turbulence calls for a revision of the uncertainty concept (Lin et al., 2020).

With regard to the red cluster, some interesting future research areas concern the drivers of SC digital capabilities, with particular attention to business model innovation processes (Alzoubi and Yanamandra, 2020; Chen, 2019; DeGroote and Marx, 2013). The literature has also paid little attention to digital technologies and competencies' role in favouring trust-based relationships between SC members' processes (Alzoubi and Yanamandra, 2020; Chen, 2019; DeGroote and Marx, 2013) and, as a consequence, in enhancing organizational agility (Dehgani and Jafari Navimipour, 2019; Ngai et al., 2011). Finally, there is a need for a deeper analysis of SC sustainability's impact on the effective use of digital technologies in production and logistic activities (Gunasekaran et al., 2019, 2018).

**Table 7.** Digitalization and agility: the main research gaps

| Research gaps | References |
|---|---|
| **Green Cluster: information technology capabilities and organizational agility** | |
| Investigating the complementarity between different digital capabilities for company agility purposes in specific competitive contexts. | Gao et al. (2020); Liu et al. (2018). |
| Expanding the meaning of environmental uncertainty and exploring its impact on digitalization-based agility more deeply. | Lin et al., (2020); Baloch et al. (2018). |
| A more in-depth exploring of how an effective IT assimilation strategy could enhance the relationship between company absorptive capacity and organizational agility. | Mao et al., (2021); Martínez-Caro et al. (2020). |
| Investigating which non-IT capabilities could complement IT competencies and digital platforms effectively to optimize organizational agility and how this could be done. | Ravichandran (2018). |
| **Red Cluster: digitalization and supply chain agility** | |
| Deepening the relationship between business model innovation and supply chain digitalization processes. | Agrawal et al., (2019); Brenner, (2018); Warner and Wager, (2019). |
| Investigating how digital technologies could favour trust-based relationship between supply chain partners. | Alzoubi and Yanamandra, (2020), Chen, (2019); Degroote and Marx (2013). |
| Learning more about which digital competences could be leveraged and integrated to obtain superior levels of supply chain agility and how this could be done. | Dehgani and Jafari Navimipour, (2019); Ngai et al., (2011). |
| Exploring how new digital technologies could enhance supply chain production's and logistic processes' sustainability. | Gunasekara et al. (2019); Gunasekara et al. (2018). |
| **Blue Cluster: big data analytics capabilities, agility and performance** | |
| Learning more about which big data analytics capabilities mostly impact a company's ambidexterity in different competitive environments and the soft aspect of BDACs' impact on organizational ambidexterity. | Fosso Wamba et al., (2020); Khanra et al. (2020b); Rialti et al., (2019); Shams et al., (2021). |
| Developing specific key performance indicators to assess how BDA-agility interconnections impact different business areas and processes' performance. | Asrini et al. (2020); Corte-Real et al. (2017). |
| Exploring the coherences that should be achieved and maintained between a company's need for BD usage and the BDA competencies required to fully exploit BDA for agility purposes | Barlette and Baillette (2020); Ghasemaghaei et al. (2017); Kisielnicki and Misiak (2016). |
| Investigating which barriers or enablers condition an effective adoption of new digital technologies for agility purposes and the effects on OA of using different combinations of digital technologies. | Akhtar et al., (2018); Pitafi et al. (2020); Rane and Narvel (2019); Sheel and Nath (2019). |

In the blue cluster, further research could investigate the drivers affecting the relationship between BDACs, ambidexterity and agility (Fosso Wamba et al., 2020; Rialti et al., 2019), as well as the specific key performance indicators to be used to assess how the BDA-agility interconnections impact different business areas' performance (Asrini et al., 2020; Corte-Real et al., 2017). Further future research should investigate how firms should therefore adapt their organizational learning processes to fully exploit BDACs (Barlette and Baillette, 2020; Ghasemaghaei et al., 2017). Finally, research on new digital technologies, such as blockchain, IoT, AI and social media, is still at an early stage and researchers should embrace the task of deepening the barriers and enablers that characterize the adoption of these technologies. This research should also determine what the best match between these technologies could be to enhance firms' agility (Akhtar et al., 2018; Pitafi et al., 2020; Rane and Narvel, 2019; Sheel and Nath, 2019).

To inspire future research directions, we next present a number of research propositions based on the above-mentioned research gaps. The propositions are not industry- or case specific, but may be relevant in different industrial or economic contexts.

*6.1 Research propositions for a future research agenda*

Based on the review of the literature presented in section 5 and by following the practices (Hughes et al., 2019), 13 unique research propositions are presented in the next sections and aiming to inspire further research on the relationships between digitalization and agility.

*6.1.1. The variable and complementary effects of the diverse digital capabilities in diverse environmental and organizational contexts*

Scholars have largely investigated the direct and indirect effects of different technical and managerial digital capabilities on company agility and performance (Gao et al., 2020; Lu and K. (Ram) Ramamurthy, 2011; Overby et al., 2006; Sambamurthy et al., 2003). Nevertheless, the importance and role of each capability depend on the characteristics of the different environmental and organizational contexts. For example, in the case of companies endowed with a high level of absorptive capacity, digital flexibility is more important than infrastructure integration capability (Gao et al., 2020). Furthermore, in times of economic recession (such as the one we are experiencing due to the current global health crisis), it could be appropriate to prioritize IT proactiveness compared to IT flexibility or integration. As a consequence, a deeper understanding of the complementary and interaction mechanisms between the different digital capabilities in the diverse environmental and organizational contexts could provide insights on how to effectively leverage these mechanisms.

**Proposition 1**. *Researchers should deepen their understanding of the complementary and interaction mechanisms between the different digital capabilities that impact organizational agility in diverse contexts.*

*6.1.2 Exploring the effects of new dimensions of environmental uncertainty on the relationship between digital capabilities and agility*

Environmental dynamism and complexity (Lin et al., 2020) are key moderating variables of the relationship between digital capabilities and organizational agility. Scholars have interpreted environmental uncertainty as mainly related to market and industry factors. However, other environmental factors, such as those related to institutional pressures, the level of government support, and the rise of global emergencies (such as the global health crisis we are currently experiencing), also affect firms' strategic growth. Future research should therefore deepen the effects of these types of uncertainty on the relationship between digital capabilities and agility.

**Proposition 2.** *There is a need to expand the meaning of environmental uncertainty for a better understanding of all of its possible impacts on the relationship between digital capabilities and agility.*

*6.1.3. Digital technology assimilation strategy as an amplifier of the absorptive capacity–agility relationship*

According to the knowledge-based view, organizational agility should be based on a superior absorptive capacity that enables effective knowledge acquisition and exploitation processes. In turn, absorptive capacity's positive effect on agility depends on an adequate digital technology assimilation attitude, i.e., the degree to which digital technologies are known and used within organizational processes (Mao et al., 2021; Martínez-Caro et al., 2020). Currently, a relevant digital infrastructure investment can become a significant amplifier of the absorptive capacity-agility relationship only when an effective digital assimilation strategy guides it.

**Proposition 3.** *Developing an effective digital technology assimilation attitude could be an effective answer to the need for ever-higher levels of agility due to the increasingly rapidly changing environment. Further studies are therefore needed to investigate the digital technology assimilation strategies that could maximize absorptive capacity's impact on organizational agility.*

*6.1.4. The role of the non–IT complementary capabilities*

Scholars have identified countless types of digital capabilities, such as technical, managerial, and relational capabilities, which have relevant and positive effects on organizational agility (Gao et al., 2020; Liu et al., 2018). Ravichandran (2018) finds that innovation capacity, i.e., a firm's propensity to innovate, is a crucial non-IT complementary capability that affects organizational agility positively. Innovation capacity is, however, a very broad concept that can be defined in many different ways.

**Proposition 4.** *Scholars should undertake a deeper investigation of the role that non-IT complementary capabilities play in the IT capabilities-agility relationship, identify and define these complementary capabilities precisely, and investigate how they could more effectively allow firms to leverage their digital platforms for agility purposes.*

*6.1.5. Supply chain digitalization and business model innovation*

New digital technologies, such as the IoT, blockchain, cloud computing and big data, allow firms to constantly evolve and reconfigure their business models and value propositions (Brenner, 2018). Nevertheless, for digital transformation to effectively renew the company's strategy and culture (Brenner, 2018; Warner and Wäger, 2019), several barriers hindering this process need to be overcome (Agrawal et al., 2019). While there have been several research studies on business model innovations based on digitalization at the firm level (Brenner, 2018), a similar investigation at the supply chain level is still lacking.

**Proposition 5.** *Researchers should undertake a deeper investigation of the digitalization strategies that could impact business model innovation effectively at the supply chain level and the barriers to be overcome to effectively develop and implement these strategies.*

*6.1.6. Leveraging digital integration strategies to enhance trust-based relationships at the SC level*
Trust is an important enabler of information-sharing processes throughout the entire SC; consequently, it is crucial to achieve coordinated responses to market changes and high levels of innovativeness (Alzoubi and Yanamandra, 2020; Chen, 2019; DeGroote and Marx, 2013). While it has been demonstrated that trust-based relationships impact SC agility and competitive advantage positively (Chen, 2019), further research is needed to investigate how and to what extent digital technologies enhance trust between SC partners and, in turn, SCA.

**Proposition 6.** *SCA depends on the effectiveness of the knowledge-sharing processes activated between SC partners. Trust is a fundamental enabler of these processes. In this connection, researchers should undertake a deeper investigation of which SC digital integration strategies, processes and practices should be designed and implemented to enhance trust-based relationships between SC partners, and how this should be done.*

*6.1.7. Integrating different digital competences for SC agility*
Being part of a supply chain provides the opportunity to collaborate with business partners and allows internal and external digital skills to be developed and integrated. Different types of digital competences, i.e., technical, operational and managerial (Dehgani and Jafari Navimipour, 2019), have been identified as crucial for SC agility's effective development (Ngai et al., 2011). However, as yet there is no deeper understanding of how each type of digital competence impacts SCA, and how these different types of competences can be effectively integrated at the SC level.

**Proposition 7.** *Researchers and practitioners should explore the different types of digital competences' (technical, operational and managerial) impact on SCA, as well as the required best practices for integrating these competences effectively at the SC level.*

*6.1.8. Leveraging new digital technologies for SC sustainability*

Manufacturing and logistics play a key role in reducing environmental impact. Scholars have already investigated many benefits and challenges related to the deployment of new digital technologies, such as big data, blockchain and IoT within manufacturing and logistic processes (Gunasekaran et al., 2018), as well as the role of these technologies in empowering SCA (Gunasekaran et al., 2019). Nevertheless, further studies are needed to investigate whether and how these technologies enhance the level of SC sustainability.

***Proposition 8.*** *Currently, sustainability is an imperative to which SCs must respond effectively in order to remain competitive. Digital technologies could help effectively face this challenge. In this connection, researchers should investigate which digital technologies and infrastructures could be leveraged to optimize SC sustainability and how this should be done.*

*6.1.9. Delving into the BDACs' effect on ambidexterity in different environments*

Studies demonstrating BDACs' significant impact on company ambidexterity and agility (Shams et al., 2021) have mainly adopted the DCs view, considering BDACs a particular category of DCs (Fosso Wamba et al., 2020; Rialti et al., 2019). Nevertheless, these studies have not delved into how the BDACs' impact varies in different market contexts and competitive environments, nor paid attention to the role of BDACs' soft characters, such as communicate as well as interactional and creative skills.

***Proposition 9.*** *Researchers and firms should deepen their understandings of which BDACs impact company ambidexterity in different competitive environments and how they do so. In addition, more research is needed to better understand the impact of BDACs' soft aspects on organizational ambidexterity.*

*6.1.10. Designing specific key performance indicators to assess how the BDA-agility interconnections impact different company departments and processes' performance*

Studies investigating the relationship between BDA, agility, and company performance use overall performance indicators, such as the market share or return on investments (Asrini et al., 2020; Côrte-Real et al., 2017). However, maximizing the positive effects arising from the BDACs-agility interconnections' optimization requires designing and using specific process-level key performance indicators (KPIs) suitable for understanding how these effects emerge in the different company departments and processes (Côrte-Real et al., 2017).

***Proposition 10.*** *Future research should aim to develop and test specific and suitable KPIs in order to assess the BDACs-agility interconnections' impact on different company areas and processes' performance.*

*6.1.11. Developing a coherent BDA organizational culture*

Fully exploiting BDA for agility purposes requires firms to evolve their organizational competencies and learning culture (Barlette and Baillette, 2020; Kisielnicki and Misiak, 2016) in order to reach and maintain an adequate fit between their needs to use BD and the competencies and knowledge that their employees actually possess (Ghasemaghaei et al., 2017). Nevertheless, more research is needed to investigate how this fit can be reached and maintained over time (Barlette and Baillette, 2020).

***Proposition 11.*** *Researchers should more deeply explore the system of coherences' qualitative and quantitative characteristics that need to be reached and maintained between a company's need to use BD and the required BDA organizational culture to fully exploit BDA for agility purposes.*

*6.1.12. Delving into the effects of integrating diverse digital technologies*

The literature analysing new digital technologies (such as IoT, AI and social media) as valuable company agility enablers is still in its infancy. First, it has not explored the effects of the different technologies' possible combinations, such as the effects of integrating blockchain technologies into IoT, AI or BD solutions (Sheel and Nath, 2019). Second, this literature has mainly focused on single firm case studies (Rane and Narvel, 2019) and industries that refer to IT-oriented environments and countries (Akhtar et al., 2018; Pitafi et al., 2020; Sheel and Nath, 2019). Finally, it has not focused on the cultural and technical barriers or on enablers in respect of these technologies' adoption, which has instead been investigated with regard to BDA (Ghasemaghaei et al., 2017).

***Proposition 12.*** *Researchers should investigate the effects of using different combinations of digital technologies on organizational agility. In addition, they should control if and how the industry context affects these technologies' impact on OA. Finally, further research should be carried out on the barriers or enablers that limit or facilitate new digital technologies' effective adoption for agility purposes.*

*6.1.13 Mutual and co-evolutionary interconnections between digital capabilities and agility capabilities: towards a two-way thinking approach*

Most of the literature that we have analysed originates from a one-direction thinking approach according to which digital capabilities are OA enablers, and not vice versa (e.g., Ahn, 2020). This literature has deeply analysed and demonstrated many facets of the role that digital capabilities, considered a set of dynamic capabilities (BDACs, abilities to leverage new digital solutions, such as AI, blockchain, and social media, etc.), can play in fostering agility (e.g., Ahn, 2020; Akhtar et al., 2019; Panda and Rath, 2017, 2016). Nevertheless, a few recent studies have also demonstrated the existence of an inverse relationship, pointing out that agility allows firms to successfully address the challenges related to rapid and unstable digital transformation and to effectively exploit emerging digital technologies, such as BDA, cloud computing, blockchain, and IoT (Vial, 2019). These studies show that agility creates the ideal condition for companies to fully exploit digitalization's transformative and value creation potential (Brenner, 2018). Some researchers have found that various obstacles to digitalization arise from the lack of an agile organization (Jermsittiparsert and Wajeetongratana, 2019; Kozarkiewicz, 2020), that OA is positively associated with and a key driver of company digital investment (Nwankpa and Merhout, 2020), while strategic agility is a critical dynamic capability for sensing, seizing and interpreting digital trends, as well as ensuring the crafting of an effective company digital mind set (Warner and Wäger, 2019).

Other studies have found that organizational flexibility (Dubey et al., 2019), agile leadership (Fachrunnisa et al., 2020), organizational speed (Barlette and Baillette, 2020) and lean management (Raut et al., 2021) profoundly affect a company's inclination to develop digital capabilities and the effectiveness with which these capabilities are designed and exploited. Moi and Cabiddu (2020) highlight how international companies, such as Spotahome, leverage their dynamic agile marketing capabilities to successfully reshape their digital processes. Finally, agile organizations have greater success with the implementation of BDA and business intelligence projects (Kisielnicki and Misiak, 2016; Kozarkiewicz, 2020). This emerging corpus of studies paves the way to a two-way thinking approach through which the mutual and co-evolutionary interconnections between digitalization and OA could be investigated and exploited.

***Proposition 13***. *Given its bidirectional nature, researchers should experiment with the adoption of a two-way thinking view to explore the relationship between digital capabilities and agility capabilities, in order investigate if and how this bidirectionality's intensity, forms and effects vary in the diverse environmental and organizational contexts, as well as to determine how the effective design of both sets of capabilities' balanced development might impact the firm's competitiveness and performance.*

*6.2. Theoretical contributions*

This study's theoretical contributions are six-fold. First, by answering RQ1, our study discovers and analyses three different, although interrelated, thematic clusters, focusing respectively on: (i) the relationship between ITCs and OA, and the ITCs-OA relationship's impact on firm competitiveness and innovation performance; (ii) the relationship between IT and digital technologies and agility at the supply chain level; and (iii) BDACs and other emerging digital technologies as crucial drivers of company agility. Besides being the first systematic investigation of the existing body of knowledge on the digitalization-agility interconnections, this thematic mapping constitutes an essential foundation for effectively identifying and addressing the subsequent research questions, the main research gaps and limitations, as well as the most promising future research avenues.

Second, in responding to RQs 2 and 3, we provide scientists and managers with 13 original research propositions on following new research pathways and developing new managerial solutions.

With regard to IT capabilities as OA enablers, this study suggests that researchers should: more deeply investigate the complementary and interaction mechanisms between the different digital capabilities that impact organizational agility in diverse contexts (Gao et al., 2020); explore the usage of new interpretative dimensions of environmental uncertainty (Lin et al., 2020); analyse the digital technology assimilation strategies (Mao et al., 2021; Martínez-Caro et al., 2020) that maximize absorptive capacity's impact on organizational agility; and study the role played by non-IT complementary capabilities (Ravichandran, 2018) in the relationship between IT capabilities and agility.

In terms of digitalization as an SC agility driver, we recommend that researchers should investigate: the best practices that allow digital transformation to impact SC business model innovation effectively (Brenner, 2018); which digital integration strategies should be adopted to fully leverage the trust mechanisms' agility effect at the SC level (Chen, 2019); the integration issues of the different types of digital competences (technical, operational and managerial; Dehgani and Jafari Navimipour, 2019) and their impact on SCA; as well as which digital technologies and infrastructures could be leveraged to optimize SC sustainability and how this should be done (Gunasekaran et al., 2019).

In terms of BDACs and digital technologies as OA drivers, this study recommends exploring which BDACs impact company ambidexterity (Shams et al., 2021) in different competitive environments and how this should be done; designing and testing specific KPIs to assess the BDACs-agility relationship's impact on different company areas and processes' performance (Côrte-Real et al., 2017); investigating the coherences to be maintained between a company's BD usage and its employees' BDA competencies (Ghasemaghaei et al., 2017); investigating the barriers to and enablers of an effective adoption of new digital technologies for agility purposes, as well as the effects of using different combinations of digital technologies in terms of firm agility (Sheel and Nath, 2019). Third, in response to RQ4, our last research proposition unveils a bidirectional connection between digital capabilities and agility, i.e., not only digital technology has a relevant and positive impact on OA, but also agility is a crucial driver of company digital evolution. In fact, several studies included in the green (Liu et al., 2013; Nwankpa and Merhout, 2020), red (Brenner, 2018; Jermsittiparsert and Wajeetongratana, 2019; Moi and Cabiddu, 2020; Warner and Wäger, 2019), and blue (Barlette and Baillette, 2020; Kisielnicki and Misiak, 2016; Kozarkiewicz, 2020) clusters show that OA itself is crucial for digital transformation processes' effective implementation and that various barriers to digital transformation are mostly related to the lack of an agile organization. Besides being a key driver of company digital investment (Nwankpa and Merhout, 2020), OA in its various forms, such as flexibility (Dubey et al., 2019), agile leadership (Fachrunnisa et al., 2020), organizational speed (Barlette and Baillette, 2020), agile absorptive capacity (Brenner, 2018), agile marketing capabilities (Moi and Cabiddu, 2020) and lean management (Raut et al., 2021), represents a critical capability for sensing and understanding new digital trends and technologies and developing an effective company digital culture (Warner and Wäger, 2019). Finally, an agile culture enabling firms to predict how new digital tools will affect current business processes and products (Jagtap and Duong, 2019; Scuotto et al., 2017) and a strong risk- and learning-oriented approach (Kane et al., 2015) are key enablers of an effective digital transformation.

Fourth, and directly linked to the previous point, by overcoming the traditional one-direction thinking view, according to which digital capabilities foster OA and not vice versa (e.g., Ahn, 2020; Akhtar et al., 2019; Panda and Rath, 2017, 2016), our review paves the way for a new corpus of studies investigating the mutual and co-evolutionary interconnections between digitalization and OA and stimulating a new managerial two-way thinking approach to effectively design the balanced development of both categories of capabilities, and to successfully manage their complex interconnections. In this regard, new theoretical approaches are, on the one hand, needed to effectively address this dual facetted relationship, while, on the other hand, it is precisely from this bidirectionality that researchers could derive new insights into digital transformation processes' design and implementation, and into required agile business models to remain competitive inturbulent environments.

Fifth, we adopt a DCs' perspective for the first time (Teece et al, 1997) to explore the existing literature on the interconnections between digitalization and OA. Our findings confirm that both OA and digital capabilities are, by their very nature, dynamic (Millar et al., 2018; Overby et al., 2006; Teece et al., 2016; Westerman et al., 2011) and that the DCs' lenses are a powerful tool to achieve a holistic and comprehensive understanding of the existing body of knowledge, the main research gaps, the understudied topics that need addressing, the under-investigated emergent issues, and the most promising future research avenues.

Sixth, with regard to OA's impact on digitalization, our findings provide new insights into the different agility capabilities required to accompany, facilitate and drive the different digital transformation phases. According to the digital transformation lifecycle framework (von Rosing and Etzel, 2020), digitalization evolves through various evolutionary phases (initial analysis, execution, and ongoing improvement). Furthermore, the portfolio of capabilities that a firm possesses must evolve progressively to maintain a dynamic and appropriate balance between the different capabilities sets in the various phases of a firm's development (Vokurka and Fliedner, 1998). Based on these considerations and our study's findings, Table 4 summarizes the agility capabilities that should be considered critical in the digital transformation lifecycle's various phases.

Table 8. Agility capabilities and the digital transformation lifecycle

| DT phase | Critical agility capabilities | Research propositions |
|---|---|---|
| Early Phase | - Pre-existent digital capabilities, skills and talent (Jagtap and Duong, 2019; Scuotto et al., 2017).<br>- Strategic agility to reconfigure the business models (Warner and Wagner, 2019; Brenner, 2018).<br>- Operational agility to introduce new digital tools at the organizational and supply chain level (Agrawal et al., 2019). | P. 5 Business model innovation and supply chain digitalization processes<br><br>P. 12 Barriers to and enablers of the adoption of new digital technologies |
| Execution Phase | - Managerial agility to integrate and complement existing capabilities with new digital technology capabilities (Gao et al., 2020) and<br>- Managerial agility to integrate and complement existing capabilities with non-IT capabilities, such as innovative capacity (Ravichandran, 2018). | P. 1 The complementarity between different digital capabilities<br>P. 4 The role of the non-IT complementary capabilities<br>P. 7 Integrating different digital competences to achieve SC agility |
| Ongoing phase | - An adequate digital technology assimilation attitude (Mao et al., 2021; Martinez-Caro et al, 2020);<br>- Managerial capabilities to improve and maintain trust-based relationships between value chain partners (Chen, 2019);<br>- Managerial capabilities to evolve the learning culture and find a fit between digital technologies usage and employers' knowledge (Barlette and Baillette, 2020). | P. 3 Digital technology assimilation strategy as an amplifier of the absorptive capacity–agility relationship<br>P. 6 The role of digital technologies in improving trust-based relationships between SC partners<br>P. 11 Exploring the BDA-organizational learning fit issue |

We hope this research will nurture further studies aimed at more deeply exploring the value creation potential of an effective co-development of digital capabilities and OA. In our opinion, meta-analysis (Jeyaraj and Dwivedi, 2020) could favour a deeper understanding of the overall value creation potential that researchers can generate by integrating the two constructs analysed in our review.

*6.3. Implications for practice*

Our results have managerial implications for designing and implementing effective business practices aimed at optimizing the mutual relationships between digitalization and OA. First, the process of digital evolution leading to superior levels of OA is long and complex, that needs to move through several stages of an evolutionary pathway. The relationships between digital capabilities and agility are also complex and bidirectional, with this bidirectionality becoming increasingly pronounced due to the exponential increase in the use of new digital tools and technologies such, as big data, cloud computing, blockchain, and IoT. In fact, while these technologies have been found to be key drivers of OA (e.g., Akhtar et al., 2019; Panda and Rath, 2017), the letter, in turn, is a crucial enabler of digital capabilities (e.g., Fachrunnisa et al., 2020; Jermsittiparsert and Wajeetongratana, 2019; Kisielnicki and Misiak, 2016; Kozarkiewicz, 2020) and of a company's digital culture (Warner and Wäger, 2019). In order to gain and maintain their competitive advantage in a globalized and digitalized world, firms should therefore focus strongly on the design and implementation of effective roadmaps aimed to successfully manage the mutual relationships and co-evolutionary interconnections between digitalization and OA.

Second, precisely by adopting DCs' managerial perspective organizations will be in a better position to effectively design the balanced development of both digital and agility capabilities, to manage the complex two-way relationships between them, as well as to effectively leverage agility to sense, seize and identify the opportunities that digital transformation's rapid and unstable character generates (Warner and Wager, 2020). By adopting the DCs' approach, firms can leverage digitalization to effectively capture, sense and interpret huge amounts of data on emerging environmental and technological trends in real time, as well as leverage agility to successfully and quickly respond to these trends.

Third, achieving and maintaining high levels of OA require the development of an organizational culture that is strongly oriented towards innovation, experimentation and exploitation of cutting-edge digital approaches and technologies (such as those relating to digital security, blockchain, social media, IoT and cloud computing). Managerial priorities should therefore be focused on attracting, training, and retaining employees who can master the most advanced digital technologies, and continuously develop their digital skills. Our findings strongly support that the effective implementation of the different digital transformation process phases requires managers and corporate staff to exhibit an adequate digital and learning-oriented attitude (Barlette and Baillette, 2020; Malekifar et al., 2014; Ravichandran, 2018). A suitable organizational climate and appropriate incentive mechanisms and policies should be designed and implemented to allow these digitally oriented and open-mind cultural and learning approaches to develop and take root.

Fourth, leveraging digital technologies effectively to enhance OA requires digitalization processes to be themselves agile and flexible in order to allow the continuous acquisition of new digital capabilities and the latter's effective integration with the existing ones. Managers should therefore not only invest in dedicated training and recruitment programmes aimed at developing and renewing their personnel's digital skills, but should also continuously engage in designing appropriate solutions and processes to effectively integrate and adequately valorise the complementarities between the existing IT infrastructural and cultural baggage and the new technologies to be continuously acquired and incorporated (Ghasemaghaei et al., 2017; Nwankpa and Merhout, 2020; Sjödin et al., 2021). IT flexibility is precisely essential when a digital transition process is ongoing. For example, cloud-based technologies could be an excellent solution to make the IT infrastructure suitably flexible; just think how much more easily a company could change its providers if it had all its servers in the Cloud rather than adopting another, no Cloud, solution. AI could also increase the flexibility of a company's digital capabilities by allowing the activation of autonomous, digital skills self-learning processes.

Fifth, our findings suggest that trust is a critical element for the effective implementation of the knowledge-sharing and co-innovation processes required to effectively leverage digital technologies in order to boost agility at the SC level and obtain the effective coordination of many different internal and external SC actors. In this regard, new digital technologies, such as blockchain solutions, seem particularly appropriate for ensuring significant and scalable processing power as well as high levels of accuracy and security. In turn, the latter are crucial enablers of monitoring and trust-building processes and collaborative innovation activities and they are therefore crucial levers for solving the lack of trust problems in all SC phases and for all SC partners. Furthermore, in a globalized and digitalized world, many actors, who may seem external to the SC context, might, in fact, significantly affect the SC value creation processes in many different ways. For example, if a social network user, who is not a customer and does not have direct relations with the company or the SC, posts a negative review, this could have a negative effect on the brand reputation.

Finally, achieving high levels of OA requires business models to be revisited and to evolve in order to follow emerging digitalization evolutionary patterns, which could also bring about a radical change. With regard to business model value propositions, the real-time big data collection and analysis processes that the use of BDA permits, might, for example, allow a shift from business models based on products (goods-dominant logic) to far more agile business models based on services (service-dominant logic). In such models, firms and their clients continuously co-create the value, rather than firms creating this value, which their clients subsequently use. Furthermore, with regard to business model channels, leveraging digital interfaces and tools might allow one or more intermediaries to be bypassed. At the same time, adopting a digital service-dominant logic could allow the business model revenue structure to transform from one based on rigid purchase-based payments to one based on recurrent lease-based payments, which would satisfy the needs of a far more clients.

*6.4. Limitations and future studies*

This study has certain limitations. First, the subjectivity characterizing the authors' interpretations and evaluations might have biased the selection of papers. We addressed this aspect by implementing a multiple human subject reading and screening of the papers (Tranfield et al., 2003). The Krippendorf's alpha coefficient was greater than 0.80, which supports our selection protocol's robustness.

A second limitation is that we only used the Scopus database during the search phase of this research. However, we cross-validated our findings with another prestigious scientific database, Web of Science and couldn't find any new relevant documents.

Third, our analysis focuses exclusively on the relationship between digitalization and OA without investigating either the interconnections with other key company capabilities (and correlated research fields) or the scientific patterns emerging across these interconnections (Shams et al., 2020). For example, our study does not delve into the processes and strategies of integrating digital capability-based decision making within risk and knowledge management, which is a relevant emerging topic in the management literature (Battisti et al., 2019; Dellermann et al., 2017).

Fourth, this study adopts a managerial viewpoint and does not explore the different technical benefits and restrictions that characterize various digital and ICT technologies, tools, and approaches (such as BDA, IOT, cloud computing, machine learning, and AI). Our analysis should therefore stimulate further research aimed at exploring the ideal usage conditions and environments for these diverse technologies in order to provide increasing levels of agility responsiveness to firms of different sizes and different industries.

Finally, our review has not investigated the different critical issues characterizing the digitalization-OA interconnections in B2B rather than in B2C contexts. We hope our research will stimulate further analyses to address this gap.

## 7. Conclusions

There is an ever-growing demand for studies aimed at exploring the consequences of the current digital revolution (Karimi and Walter, 2015; Vial, 2019), at investigating the value offered by each specific analytical tool to management practices (Khanra et al., 2020b) and at concretely helping managers successfully address the co-evolutionary relationship between OA and digitalization. This research meets this need by, for the first time, offering a systematic review of the literature on the digitalization-OA interconnections.

We find that both OA and digital capabilities are, by their very nature, dynamic (Teece et al., 2016; Westerman et al., 2011) and that the DCs' lenses are a powerful tool to achieve a holistic understanding of the existing body of knowledge, the main research gaps, and the most promising future research avenues.

We discover and analyse three different, although interrelated, thematic clusters, respectively focusing on big-data analytic capabilities as crucial drivers of OA, the relationship between digitalization and agility at a supply chain level, and the role of information technology capabilities in improving OA.

This study provides scientists and managers with 13 unique research propositions on relevant theoretical and practical issues (Table 9).

Table 9. A summary of research propositions

| IT capabilities as OA enablers | Digitalization as SC agility driver | BDAC and digital technologies as OA drivers |
|---|---|---|
| **Proposition 1**. Researchers should deepen their understanding of the complementary and interaction mechanisms between the different digital capabilities that impact organizational agility in diverse contexts.<br>**Proposition 2.** There is a need to expand the meaning of environmental uncertainty for a better understanding of all of its possible impacts on the relationship between digital capabilities and agility.<br>**Proposition 3.** Developing an effective digital technology assimilation attitude could be an effective answer to the need for ever-higher levels of agility due to the increasingly rapidly changing environment. Further studies are therefore needed to investigate the digital technology assimilation strategies that could maximize absorptive capacity's impact on organizational agility.<br>**Proposition 4.** Scholars should undertake a deeper investigation of the role that non-IT complementary capabilities play in the IT capabilities-agility relationship, identify and define these complementary capabilities precisely, and investigate how they could more effectively allow firms to leverage their digital platforms for agility purposes.<br>**Proposition 5.** Researchers should undertake a deeper investigation of the digitalization strategies that could impact business model innovation effectively at the supply chain level and the barriers to be overcome to effectively develop and implement these strategies. | **Proposition 6.** SCA depends on the effectiveness of the knowledge-sharing processes activated between SC partners. Trust is a fundamental enabler of these processes. In this connection, researchers should undertake a deeper investigation of which SC digital integration strategies, processes and practices should be designed and implemented to enhance trust-based relationships between SC partners, and how this should be done.<br>**Proposition 7.** Researchers and practitioners should explore the different types of digital competences' (technical, operational and managerial) impact on SCA, as well as the required best practices for integrating these competences effectively at the SC level.<br>**Proposition 8.** Currently, sustainability is an imperative to which SCs must respond effectively in order to remain competitive. Digital technologies could help effectively face this challenge. In this connection, researchers should investigate which digital technologies and infrastructures could be leveraged to optimize SC sustainability and how this should be done.<br>**Proposition 9.** Researchers and firms should deepen their understandings of which BDACs impact company ambidexterity in different competitive environments and how they do so. In addition, more research is needed to better understand the impact of BDACs' soft aspects on organizational ambidexterity. | **Proposition 10.** Future research should aim to develop and test specific and suitable KPIs in order to assess the BDACs-agility interconnections' impact on different company areas and processes' performance.<br>**Proposition 11.** Researchers should more deeply explore the system of coherences' qualitative and quantitative characteristics that need to be reached and maintained between a company's need to use BD and the required BDA organizational culture to fully exploit BDA for agility purposes.<br>**Proposition 12.** Researchers should investigate the effects of using different combinations of digital technologies on organizational agility. In addition, they should control if and how the industry context affects these technologies' impact on OA. Finally, further research should be carried out on the barriers or enablers that limit or facilitate new digital technologies' effective adoption for agility purposes.<br>**Proposition 13**. Given its bidirectional nature, researchers should experiment with the adoption of a two-way thinking view to explore the relationship between digital capabilities and agility capabilities, in order investigate if and how this bidirectionality's intensity, forms and effects vary in the diverse environmental and organizational contexts, as well as to determine how the effective design of both sets of capabilities' balanced development might impact the firm's competitiveness and performance. |

These propositions unveil several promising future research avenues. With regard to IT capabilities as OA enablers this study suggests researchers:

- more deeply investigate the complementary and interaction mechanisms between the different digital capabilities that impact organizational agility in diverse contexts (Gao et al., 2020);

- explore the usage of new interpretative dimensions of environmental uncertainty (Lin et al., 2020) for a better understanding of all of its possible impacts on the relationship between digital capabilities and agility;

explore the digital technology assimilation strategies (Mao et al., 2021) that could maximize absorptive capacity's impact on organizational agility;

- investigate the role played by non-IT complementary capabilities (Ravichandran, 2018) in the relationship between IT capabilities and agility precisely and understand how these complementary capabilities could more effectively allow firms to leverage their digital platforms for agility purposes.

In terms of digitalization as an SC agility driver (propositions 5 to 8), we recommend that researchers should:

- explore the digitalization strategies that could impact business model innovation at the supply chain
level (Brenner, 2018) and the barriers to be overcome to effectively develop and implement these strategies;

- investigate which SC digital integration strategies, processes and practices should be designed and implemented to fully leverage the trust mechanisms' agility effect at the SC level (Chen, 2019);

- investigate the different types of digital competencies' (technical, operational and managerial; Dehgani and Jafari Navimipour, 2019) impact on SCA, as well as the best practices for integrating these competencies effectively at the SC level;

- investigate which digital technologies and infrastructures could be leveraged to optimize SC sustainability and how this should be done (Gunasekaran et al., 2019).

In terms of BDACs and digital technologies as OA drivers (propositions 9 to 12), this study recommends:

- exploring which BDACs impact company ambidexterity (Shams et al., 2021) in different competitive environments;

- developing and testing specific KPIs to assess the BDACs-agility relationship's impact on different company areas and processes' performance (Côrte-Real et al., 2017);

- exploring the coherences to be maintained between a company's BD usage and its employees' BDA competencies and culture (Ghasemaghaei et al., 2017) to fully exploit BDA for agility purposes;

- delving into the barriers to and enablers of an effective adoption of new digital technologies for agility purposes, as well as the effects of using different combinations of digital technologies for agility purposes (Sheel and Nath, 2019)

Finally, the last proposition (number 13) encourages researchers to experiment with a two-way thinking view to explore the relationship between digital capabilities and agility capabilities. This approach could allow researchers to determine how the effective design of a balanced development of both sets of capabilities might impact the firm's competitiveness and performance, as well as to understand if and how the forms and effects of the bidirectional nature of this relationship vary in the diverse environmental and organizational contexts. By overcoming the traditional one-direction thinking view, according to which digital capabilities foster OA and not vice versa (e.g., Akhtar et al., 2019; Panda and Rath, 2017), our review paves the way for a new corpus of studies investigating the mutual and co-evolutionary interconnections between digitalization and OA. Although mainly focused on the relationships between agility and digital capabilities, approaches and culture, most of the propositions developed in this study could also stimulate further research on digitalization and agility's effects on corporate strategy and innovation (Ciampi et al., 2020).

**References:**

Accenture, 2020. Business Agility Report: Responding to Disruption (No. 3), Business AgilityReport. Business Agility Institute, Irvine, California.

Agrawal, P., Narain, R., Ullah, I., 2019. Analysis of barriers in implementation of digital transformation of supply chain using interpretive structural modelling approach. JM2 15, 297–317. https://doi.org/10.1108/JM2-03-2019-0066

Ahn, S.-J., 2020. Three characteristics of technology competition by IoT-driven digitization. Technological Forecasting and Social Change 157, 120062. https://doi.org/10.1016/j.techfore.2020.120062

Akhtar, P., Khan, Z., Tarba, S., Jayawickrama, U., 2018. The Internet of Things, dynamic data and information processing capabilities, and operational agility. Technological Forecasting and Social Change 136, 307–316. https://doi.org/10.1016/j.techfore.2017.04.023

Akter, S., Bandara, R., Hani, U., Fosso Wamba, S., Foropon, C., Papadopoulos, T., 2019. Analytics-based decision-making for service systems: A qualitative study and agenda for future research. International Journal of Information Management 48, 85–95. https://doi.org/10.1016/j.ijinfomgt.2019.01.020

Akter, S., Wamba, S.F., 2016. Big data analytics in E-commerce: a systematic review and agenda forfuture research. Electronic Markets 26, 173–194. https://doi.org/10.1007/s12525-016-0219-0 Alzoubi, H.M., Yanamandra, R., 2020. Investigating the mediating role of information sharingstrategy on agile supply chain. 10.5267/j.uscm 273–284. https://doi.org/10.5267/j.uscm.2019.12.004

Asrini, A., Musnaini, M., Setyawati, Y., Kumalawati, L., Fajariyah, N.A., 2020. Predictors of Firm Performance and Supply Chain: Evidence from Indonesian Pharmaceuticals Industry. International Journal of Supply Chain Management 9, 1080–1087.

Ayabakan, S., Bardhan, I.R., Zheng, Z., 2017. A data envelopment analysis approach to estimate it- enabled production capability. Mis Quarterly 41.

Baloch, M.A., Meng, F., Bari, M.W., 2018. Moderated mediation between IT capability and organizational agility. HSM 37, 195–206. https://doi.org/10.3233/HSM-17150

Bargshady, G., Zahraee, S.M., Ahmadi, M., Parto, A., 2016. The effect of information technology onthe agility of the supply chain in the Iranian power plant industry. Journal of Manufacturing Technology Management 27, 427–442. https://doi.org/10.1108/JMTM-11-2015-0093

Barlette, Y., Baillette, P., 2020. Big data analytics in turbulent contexts: towards organizational change for enhanced agility. Production Planning & Control 1–18. https://doi.org/10.1080/09537287.2020.1810755

Battisti, E., Shams, S.M.R., Sakka, G., Miglietta, N., 2019. Big data and risk management in business processes: implications for corporate real estate. Business Process Management Journal 26, 1141–1155. https://doi.org/10.1108/BPMJ-03-2019-0125

Behera, R.K., Bala, P.K., Dhir, A., 2019. The emerging role of cognitive computing in healthcare: asystematic literature review. International journal of medical informatics 129, 154–166.

Bessant, J., Francis, D., Meredith, S., Kaplinsky, R., Brown, S., 2001. Developing manufacturing agility in SMEs. International Journal of Technology Management 22, 28–28.

Bhatt, Y., Ghuman, K., Dhir, A., 2020. Sustainable manufacturing. Bibliometrics and content analysis. Journal of Cleaner Production 260, 120988.

Boyack, K.W., Klavans, R., 2010. Co-citation analysis, bibliographic coupling, and direct citation: Which citation approach represents the research front most accurately? Journal of the American Society for Information Science and Technology 61, 2389–2404. https://doi.org/10.1002/asi.21419

Brenner, B., 2018. Transformative Sustainable Business Models in the Light of the Digital Imperative—A Global Business Economics Perspective. Sustainability 10, 4428. https://doi.org/10.3390/su10124428

Bresciani, S., Ciampi, F., Meli, F., Ferraris, A., 2021. Using big data for co-innovation processes: Mapping the field of data-driven innovation, proposing theoretical developments and providing a research agenda. International Journal of Information Management 102347. https://doi.org/10.1016/j.ijinfomgt.2021.102347

Breu, K., Hemingway, C.J., Strathern, M., Bridger, D., 2002. Workforce Agility: The New Employee Strategy for the Knowledge Economy. Journal of Information Technology 17, 21–31. https://doi.org/10.1080/02683960110132070

Broadbent, M., Weill, P., Neo, B.S., 1999. Strategic context and patterns of IT infrastructure capability. Journal of Strategic Information Systems 8, 157–187. https://doi.org/10.1016/S0963-8687(99)00022-0

Cai, Z., Liu, H., Huang, Q., Liang, L., 2019. Developing organizational agility in product innovation: the roles of IT capability, KM capability, and innovative climate: Developing organizational agility in product innovation. R&D Management 49, 421–438. https://doi.org/10.1111/radm.12305

Calatayud, A., Mangan, J., Christopher, M., 2019. The self-thinking supply chain. SCM 24, 22–38. https://doi.org/10.1108/SCM-03-2018-0136

Cao, Q., Dowlatshahi, S., 2005. The impact of alignment between virtual enterprise and information technology on business performance in an agile manufacturing environment. Journal of Operations Management 23, 531–550. https://doi.org/10.1016/j.jom.2004.10.010

Cepeda, J., Arias-Pérez, J., 2019. Information technology capabilities and organizational agility: The mediating effects of open innovation capabilities. MBR 27, 198–216. https://doi.org/10.1108/MBR-11-2017-0088

Chakravarty, A., Grewal, R., Sambamurthy, V., 2013. Information Technology Competencies, Organizational Agility, and Firm Performance: Enabling and Facilitating Roles. Information Systems Research 24, 976–997. https://doi.org/10.1287/isre.2013.0500

Chan, C.M.L., Teoh, S.Y., Yeow, A., Pan, G., 2019. Agility in responding to disruptive digital innovation: Case study of an SME. Information Systems Journal 29, 436–455. https://doi.org/10.1111/isj.12215

Chen, C.-J., 2019. Developing a model for supply chain agility and innovativeness to enhance firms' competitive advantage. MD 57, 1511–1534. https://doi.org/10.1108/MD-12-2017-1236

Choudhury, A., Behl, A., Sheorey, P.A., Pal, A., 2021. Digital supply chain to unlock new agility: a TISM approach. Benchmarking: An International Journal.

Christopher, M., Ryals, L.J., 2014. The Supply Chain Becomes the Demand Chain. J Bus Logist 35, 29–35. https://doi.org/10.1111/jbl.12037

Ciampi, F., Demi, S., Magrini, A., Marzi, G., Papa, A., 2021. Exploring the impact of big data analytics capabilities on business model innovation: The mediating role of entrepreneurial orientation. Journal of Business Research 123, 1–13. https://doi.org/10.1016/j.jbusres.2020.09.023

Ciampi, F., Marzi, G., Demi, S., Faraoni, M., 2020. The big data-business strategy interconnection: a grand challenge for knowledge management. A review and future perspectives. Journal of Knowledge Management 24, 1157–1176. https://doi.org/10.1108/JKM-02-2020-0156

Côrte-Real, N., Oliveira, T., Ruivo, P., 2017. Assessing business value of Big Data Analytics in European firms. Journal of Business Research 70, 379–390. https://doi.org/10.1016/j.jbusres.2016.08.011

Crocitto, M., Youssef, M., 2003. The human side of organizational agility. Industr Mngmnt & Data Systems 103, 388–397. https://doi.org/10.1108/02635570310479963

Cunha, M.P. e, Gomes, E., Mellahi, K., Miner, A.S., Rego, A., 2020. Strategic agility through improvisational capabilities: Implications for a paradox-sensitive HRM. Human Resource Management Review 30, 100695. https://doi.org/10.1016/j.hrmr.2019.100695

DeGroote, S.E., Marx, T.G., 2013. The impact of IT on supply chain agility and firm performance: An empirical investigation. International Journal of Information Management 33, 909–916. https://doi.org/10.1016/j.ijinfomgt.2013.09.001

Dehgani, R., Jafari Navimipour, N., 2019. The impact of information technology and communication systems on the agility of supply chain management systems. K 48, 2217–2236. https://doi.org/10.1108/K-10-2018-0532

Del Giudice, M., 2016. Discovering the Internet of Things (IoT) within the business process management: a literature review on technological revitalization. Business Process Management Journal 22. https://doi.org/10.1108/BPMJ-12-2015-0173

Delgado García, J.B., De Quevedo Puente, E., Blanco Mazagatos, V., 2015. How Affect Relates to Entrepreneurship: A Systematic Review of the Literature and Research Agenda. International Journal of Management Reviews 17, 191–211. https://doi.org/10.1111/ijmr.12058

Dellermann, D., Fliaster, A., Kolloch, M., 2017. Innovation risk in digital business models: the German energy sector. Journal of Business Strategy 38, 35–43. https://doi.org/10.1108/JBS-07-2016-0078

Di Vaio, A., Palladino, R., Pezzi, A., Kalisz, D.E., 2021. The role of digital innovation in knowledge management systems: A systematic literature review. Journal of Business Research 123, 220–231. https://doi.org/10.1016/j.jbusres.2020.09.042

Dijkman, R.M., Sprenkels, B., Peeters, T., Janssen, A., 2015. Business models for the Internet of Things. International Journal of Information Management 35, 672–678. https://doi.org/10.1016/j.ijinfomgt.2015.07.008

Doz, Y., 2020. Fostering strategic agility: How individual executives and human resource practices contribute. Human Resource Management Review 30, 100693. https://doi.org/10.1016/j.hrmr.2019.100693

Dubey, R., Gunasekaran, A., Childe, S.J., 2019. Big data analytics capability in supply chain agility: The moderating effect of organizational flexibility. MD 57, 2092–2112. https://doi.org/10.1108/MD-01-2018-0119

E. Prescott, M., 2014. Big data and competitive advantage at Nielsen. Management Decision 52, 573–601. https://doi.org/10.1108/MD-09-2013-0437

Fachrunnisa, O., Adhiatma, A., Majid, M.N.A., Lukman, N., 2020. Towards SMEs' digital transformation: The role of agile leadership and strategic flexibility. JSBS 30, 65–85.

Falagas, M.E., Pitsouni, E.I., Malietzis, G., Pappas, G., 2008. Comparison of PubMed, Scopus, web of science, and Google scholar: strengths and weaknesses. The FASEB Journal 22, 338–342.

Felipe, C.M., Leidner, D.E., Roldán, J.L., Leal- Rodríguez, A.L., 2020. Impact of IS Capabilities on Firm Performance: The Roles of Organizational Agility and Industry Technology Intensity. Decision Sciences 51, 575–619. https://doi.org/10.1111/deci.12379

Fitzgerald, M., Kruschwitz, N., Bonnet, D., Welch, M., 2014. Embracing Digital Technology: a new strategic imperative. 55, 1–16.

Fosso Wamba, S., Dubey, R., Gunasekaran, A., Akter, S., 2020. The performance effects of big data analytics and supply chain ambidexterity: The moderating effect of environmental dynamism.

International Journal of Production Economics 222, 107498. https://doi.org/10.1016/j.ijpe.2019.09.019

Gao, P., Zhang, J., Gong, Y., Li, H., 2020. Effects of technical IT capabilities on organizational agility: The moderating role of IT business spanning capability. IMDS 120, 941–961. https://doi.org/10.1108/IMDS-08-2019-0433

Garcia-Alcaraz, J.L., Maldonado-Macias, A.A., Alor-Hernandez, G., Sanchez-Ramirez, C., 2017. The impact of information and communication technologies (ICT) on agility, operating, and economical performance of supply chain. Adv produc engineer manag 12, 29–40. https://doi.org/10.14743/apem2017.1.237

Gaur, A., Kumar, M., 2018. A systematic approach to conducting review studies: An assessment of content analysis in 25 years of IB research. Journal of World Business 53, 280–289. https://doi.org/10.1016/j.jwb.2017.11.003

Ghasemaghaei, M., Hassanein, K., Turel, O., 2017. Increasing firm agility through the use of data analytics: The role of fit. Decision Support Systems 101, 95–105. https://doi.org/10.1016/j.dss.2017.06.004

Giannakis, M., Louis, M., 2016. A multi-agent based system with big data processing for enhanced supply chain agility. JEIM 29, 706–727. https://doi.org/10.1108/JEIM-06-2015-0050

Glaser, B.G., Strauss, A.L., Strutzel, E., 1968. The discovery of grounded theory; strategies for qualitative research. Nursing research 17, 364.

Glenn, M., 2009. Organizational agility: how business can survive and thrive in turbulent times. Economist Intelligence Unit Limited, London.

Gligor, D., Gligor, N., Holcomb, M., Bozkurt, S., 2019. Distinguishing between the concepts of supply chain agility and resilience: A multidisciplinary literature review. The International Journal of Logistics Management 30, 467–487. https://doi.org/10.1108/IJLM-10-2017-0259

Gligor, D.M., Holcomb, M.C., 2012. Understanding the role of logistics capabilities in achieving supply chain agility: a systematic literature review. Supply Chain Management: An International Journal 17, 438–453. https://doi.org/10.1108/13598541211246594

Gunasekaran, A., 1999. Agile manufacturing: A framework for research and development. International Journal of Production Economics 62, 87–105. https://doi.org/10.1016/S0925-5273(98)00222-9

Gunasekaran, A., Yusuf, Y.Y., Adeleye, E.O., Papadopoulos, T., 2018. Agile manufacturing practices: the role of big data and business analytics with multiple case studies. International Journal of Production Research 56, 385–397. https://doi.org/10.1080/00207543.2017.1395488

Gunasekaran, A., Yusuf, Y.Y., Adeleye, E.O., Papadopoulos, T., Kovvuri, D., Geyi, D.G., 2019. Agile manufacturing: an evolutionary review of practices. International Journal of Production Research 57, 5154–5174. https://doi.org/10.1080/00207543.2018.1530478

Günther, W.A., Rezazade Mehrizi, M.H., Huysman, M., Feldberg, F., 2017. Debating big data: A literature review on realizing value from big data. The Journal of Strategic Information Systems 26, 191–209. https://doi.org/10.1016/j.jsis.2017.07.003

Hajli, N., Tajvidi, M., Gbadamosi, A., Nadeem, W., 2020. Understanding market agility for new product success with big data analytics. Industrial Marketing Management 86, 135–143. https://doi.org/10.1016/j.indmarman.2019.09.010

Hatzijordanou, N., Bohn, N., Terzidis, O., 2019. A systematic literature review on competitor analysis: status quo and start-up specifics. Management Review Quarterly 69, 415–458.

Hess, T., Matt, C., Benlian, A., Wiesböck, F., 2016. Options for Formulating a Digital Transformation Strategy. MIS Quarterly Executive 123–139.

Huang, Y., Singh, P.V., Ghose, A., 2015. A Structural Model of Employee Behavioral Dynamics in Enterprise Social Media. Management Science 61, 2825–2844.

Hughes, L., Dwivedi, Y.K., Misra, S.K., Rana, N.P., Raghavan, V., Akella, V., 2019. Blockchain research, practice and policy: Applications, benefits, limitations, emerging research themes and research agenda. International Journal of Information Management 49, 114–129. https://doi.org/10.1016/j.ijinfomgt.2019.02.005

Inman, R.A., Sale, R.S., Green, K.W., Whitten, D., 2011. Agile manufacturing: Relation to JIT, operational performance and firm performance. Journal of Operations Management 29, 343–355. https://doi.org/10.1016/j.jom.2010.06.001

Isensee, C., Teuteberg, F., Griese, K.-M., Topi, C., 2020. The relationship between organizational culture, sustainability, and digitalization in SMEs: A systematic review. Journal of Cleaner Production 275, 122944. https://doi.org/10.1016/j.jclepro.2020.122944

Jagtap, S., Duong, L.N.K., 2019. Improving the new product development using big data: a case study of a food company. British Food Journal 121, 2835–2848. https://doi.org/10.1108/BFJ-02-2019-0097

Jermsittiparsert, K., Wajeetongratana, P., 2019. The Role of Organizational Culture and It Competency in Determining the Supply Chain Agility in the Small and Medium-Size Enterprises. International Journal of Innovation 5, 17.

Jeyaraj, A., Dwivedi, Y.K., 2020. Meta-analysis in information systems research: Review and recommendations. International Journal of Information Management 55, 102226. https://doi.org/10.1016/j.ijinfomgt.2020.102226

Kane, G.C., 2015. Enterprise Social Media: Current Capabilities and Future Possibilities. MIS Quarterly Executive 14, 1–16.

Kane, G.C., Palmer, D., Phillips, A.N., Kiron, D., Buckley, N., 2015. Strategy, not Technology, Drives Digital Transformation. MIT Sloan Management Review and Deloitte University Press.

Karimi, J., Walter, Z., 2015. The Role of Dynamic Capabilities in Responding to Digital Disruption: A Factor-Based Study of the Newspaper Industry. Journal of Management Information Systems 32, 39–81. https://doi.org/10.1080/07421222.2015.1029380

Kaur, P., Dhir, A., Tandon, A., Alzeiby, E.A., Abohassan, A.A., 2020. A systematic literature review on cyberstalking. An analysis of past achievements and future promises. Technological Forecasting and Social Change 120426.

Khan, A., Krishnan, S., Dhir, A., 2021. Electronic government and corruption: Systematic literature review, framework, and agenda for future research. Technological Forecasting and Social Change 167, 120737. https://doi.org/10.1016/j.techfore.2021.120737

Khanra, S., Dhir, A., Islam, A.K.M.N., Mäntymäki, M., 2020a. Big data analytics in healthcare: a systematic literature review. Enterprise Information Systems 14, 878–912. https://doi.org/10.1080/17517575.2020.1812005

Khanra, S., Dhir, A., Kaur, P., Mäntymäki, M., 2021a. Bibliometric analysis and literature review of ecotourism: Toward sustainable development. Tourism Management Perspectives 37, 100777.

Khanra, S., Dhir, A., Mäntymäki, M., 2020b. Big data analytics and enterprises: a bibliometric synthesis of the literature. Enterprise Information Systems 14, 737–768. https://doi.org/10.1080/17517575.2020.1734241

Khanra, S., Dhir, A., Parida, V., Kohtamäki, M., 2021b. Servitization research: A review and bibliometric analysis of past achievements and future promises. Journal of Business Research 131, 151–166.

Khin, S., Ho, T.C., 2018. Digital technology, digital capability and organizational performance: A mediating role of digital innovation. International Journal of Innovation Science 11, 177–195. https://doi.org/10.1108/IJIS-08-2018-0083

Kisielnicki, J.A., Misiak, A.M., 2016. Effectiveness of Agile Implementation Methods in Business Intelligence Projects from an End-user Perspective. InformingSciJ 19, 161–172. https://doi.org/10.28945/3515

Kozarkiewicz, A., 2020. General and Specific: The Impact of Digital Transformation on Project Processes and Management Methods. Foundations of Management 12, 237–248. https://doi.org/10.2478/fman-2020-0018

Kraus, S., Breier, M., Dasí-Rodríguez, S., 2020. The art of crafting a systematic literature review in entrepreneurship research. International Entrepreneurship and Management Journal 16, 1023–1042. https://doi.org/10.1007/s11365-020-00635-4

Kuusisto, M., 2017. Organizational effects of digitalization: A literature review. International Journal of Organization Theory and Behavior 20, 341–362. https://doi.org/10.1108/IJOTB-20-03-2017-B003

Lee, J., Cho, H., Kim, Y.S., 2015. Assessing business impacts of agility criterion and order allocation strategy in multi-criteria supplier selection. Expert Systems with Applications 42, 1136–1148. https://doi.org/10.1016/j.eswa.2014.08.041

Legner, C., Eymann, T., Hess, T., Matt, C., Böhmann, T., Drews, P., Mädche, A., Urbach, N., Ahlemann, F., 2017. Digitalization: Opportunity and Challenge for the Business and Information Systems Engineering Community. Bus Inf Syst Eng 59, 301–308. https://doi.org/10.1007/s12599-017-0484-2

Lenka, S., Parida, V., Wincent, J., 2017. Digitalization capabilities as enablers of value co-creation in servitizing firms. Psychology & marketing 34, 92–100.

Levallet, N., Chan, Y.E., 2018. Role of Digital Capabilities in Unleashing the Power of Managerial Improvisation. MIS Quarterly Executive 17.

Li, L., Su, F., Zhang, W., Mao, J.-Y., 2018. Digital transformation by SME entrepreneurs: A capability perspective. Information Systems Journal 28, 1129–1157. https://doi.org/10.1111/isj.12153

Lin, C.-T., Chiu, H., Tseng, Y.-H., 2006. Agility evaluation using fuzzy logic. International Journal of Production Economics 101, 353–368. https://doi.org/10.1016/j.ijpe.2005.01.011

Lin, J., Li, L., Luo, X. (Robert), Benitez, J., 2020. How do agribusinesses thrive through complexity? The pivotal role of e-commerce capability and business agility. Decision Support Systems 135, 113342. https://doi.org/10.1016/j.dss.2020.113342

Liu, H., Ke, W., Wei, K.K., Hua, Z., 2013. The impact of IT capabilities on firm performance: The mediating roles of absorptive capacity and supply chain agility. Decision Support Systems 54, 1452–1462. https://doi.org/10.1016/j.dss.2012.12.016

Liu, S., Chan, F.T.S., Yang, J., Niu, B., 2018. Understanding the effect of cloud computing on organizational agility: An empirical examination. International Journal of Information Management 43, 98–111. https://doi.org/10.1016/j.ijinfomgt.2018.07.010

Lu, Y., K. (Ram) Ramamurthy, 2011. Understanding the Link Between Information Technology Capability and Organizational Agility: An Empirical Examination. MIS Quarterly 35, 931. https://doi.org/10.2307/41409967

Lucas, H.C., Goh, J.M., 2009. Disruptive technology: How Kodak missed the digital photography revolution. The Journal of Strategic Information Systems 18, 46–55. https://doi.org/10.1016/j.jsis.2009.01.002

Luqman, A., Talwar, S., Masood, A., Dhir, A., 2021. Does enterprise social media use promote employee creativity and well-being? Journal of Business Research 131, 40–54. https://doi.org/10.1016/j.jbusres.2021.03.051

Mak, H.-Y., Shen, Z.-J.M., 2021. When Triple-A Supply Chains Meet Digitalization: The Case of JD.com's C2M Model. Production and Operations Management 30, 656–665. https://doi.org/10.1111/poms.13307

Malekifar, S., Taghizadeh, S.K., Rahman, S.A., Khan, S.U.R., 2014. Organizational Culture, IT Competence, and Supply Chain Agility in Small and Medium-Size Enterprises. Global Business and Organizational Excellence 33, 69–75. https://doi.org/10.1002/joe.21574

Mandal, S., 2018. An examination of the importance of big data analytics in supply chain agility development: A dynamic capability perspective. MRR 41, 1201–1219. https://doi.org/10.1108/MRR-11-2017-0400

Mao, H., Liu, S., Zhang, J., 2015. How the effects of IT and knowledge capability on organizational agility are contingent on environmental uncertainty and information intensity. Information Development 31, 358–382. https://doi.org/10.1177/0266666913518059

Mao, H., Liu, S., Zhang, J., Zhang, Y., Gong, Y., 2021. Information technology competency and organizational agility: roles of absorptive capacity and information intensity. ITP 34, 421–451. https://doi.org/10.1108/ITP-12-2018-0560

Marhraoui, M.A., El Manouar, A., 2017. IT innovation and firm's sustainable performance: The mediating role of organizational agility, in: Proceedings of the 9th International Conference on Information Management and Engineering, ICIME 2017. Association for Computing Machinery, New York, NY, USA, pp. 150–156. https://doi.org/10.1145/3149572.3149578

Martínez-Caro, E., Cepeda-Carrión, G., Cegarra-Navarro, J.G., Garcia-Perez, A., 2020. The effect of information technology assimilation on firm performance in B2B scenarios. IMDS 120, 2269–2296. https://doi.org/10.1108/IMDS-10-2019-0554

Marzi, G., Ciampi, F., Dalli, D., Dabic, M., 2020. New product development during the last ten years: The ongoing debate and future avenues. IEEE Transactions on Engineering Management. https://doi.org/10.1109/TEM.2020.2997386

McAfee, A., Brynjolfsson, E., 2012. Big Data: The Management Revolution. Harward Business Review 6, 1–9.

Mehta, N., Pandit, A., 2018. Concurrence of big data analytics and healthcare: A systematic review. International journal of medical informatics 114, 57–65.

Melián-Alzola, L., Fernández-Monroy, M., Hidalgo-Peñate, M., 2020. Information technology capability and organisational agility: A study in the Canary Islands hotel industry. Tourism Management Perspectives 33, 100606. https://doi.org/10.1016/j.tmp.2019.100606

Mikalef, P., Pateli, A., 2017. Information technology-enabled dynamic capabilities and their indirect effect on competitive performance: Findings from PLS-SEM and fsQCA. Journal of Business Research 70, 1–16. https://doi.org/10.1016/j.jbusres.2016.09.004

Millar, C.C.J.M., Groth, O., Mahon, J.F., 2018. Management Innovation in a VUCA World: Challenges and Recommendations. California Management Review 61, 5–14. https://doi.org/10.1177/0008125618805111

Moi, L., Cabiddu, F., 2020. Leading digital transformation through an Agile Marketing Capability: the case of Spotahome. J Manag Gov. https://doi.org/10.1007/s10997-020-09534-w

Muduli, A., 2013. Workforce Agility: A Review of Literature. IUP Journal of Management Research 12, 55–65.

Nadkarni, S., Prügl, R., 2021. Digital transformation: a review, synthesis and opportunities for future research. Manag Rev Q 71, 233–341. https://doi.org/10.1007/s11301-020-00185-7

Nandi, S., Sarkis, J., Hervani, A., Helms, M., 2020. Do blockchain and circular economy practices improve post COVID-19 supply chains? A resource-based and resource dependence perspective. Industrial Management & Data Systems 121, 333–363. https://doi.org/10.1108/IMDS-09-2020-0560

Ngai, E.W.T., Chau, D.C.K., Chan, T.L.A., 2011. Information technology, operational, and management competencies for supply chain agility: Findings from case studies. The Journal of Strategic Information Systems 20, 232–249. https://doi.org/10.1016/j.jsis.2010.11.002

Nusrat, A., He, Y., Luqman, A., Waheed, A., Dhir, A., 2021. Enterprise social media and cyber-slacking: A Kahn's model perspective. Information & Management 58, 103405. https://doi.org/10.1016/j.im.2020.103405

Nwankpa, J.K., Merhout, J.W., 2020. Exploring the Effect of Digital Investment on IT Innovation. Sustainability 12, 7374. https://doi.org/10.3390/su12187374

Osei, C., Amankwah-Amoah, J., Khan, Z., Omar, M., Gutu, M., 2018. Developing and deploying marketing agility in an emerging economy: the case of Blue Skies. International Marketing Review 36, 190–212. https://doi.org/10.1108/IMR-12-2017-0261

Overby, E., Bharadwaj, A., Sambamurthy, V., 2006. Enterprise agility and the enabling role of information technology. null 15, 120–131. https://doi.org/10.1057/palgrave.ejis.3000600

Panda, S., 2021. Strategic IT-business alignment capability and organizational performance: roles of organizational agility and environmental factors. Journal of Asia Business Studies ahead-of-print. https://doi.org/10.1108/JABS-09-2020-0371

Panda, S., Rath, S.K., 2017. The effect of human IT capability on organizational agility: an empirical analysis. MRR 40, 800–820. https://doi.org/10.1108/MRR-07-2016-0172

Panda, S., Rath, S.K., 2016. Investigating the structural linkage between IT capability and organizational agility: A study on Indian financial enterprises. JEIM 29, 751–773. https://doi.org/10.1108/JEIM-04-2015-0033

Paschou, T., Rapaccini, M., Adrodegari, F., Saccani, N., 2020. Digital servitization in manufacturing: A systematic literature review and research agenda. Industrial Marketing Management 89, 278–292. https://doi.org/10.1016/j.indmarman.2020.02.012

Patil, M., Suresh, M., 2019. Modelling the Enablers of Workforce Agility in IoT Projects: A TISM Approach. Glob J Flex Syst Manag 20, 157–175. https://doi.org/10.1007/s40171-019-00208-7

Paulraj, A., Chen, I.J., 2007. Strategic Buyer–Supplier Relationships, Information Technology and External Logistics Integration. Journal of Supply Chain Management 43, 2–14. https://doi.org/10.1111/j.1745-493X.2007.00027.x

Pellegrini, M.M., Ciampi, F., Marzi, G., Orlando, B., 2020. The relationship between knowledge management and leadership: mapping the field and providing future research avenues. Journal of Knowledge Management 24, 1445–1492. https://doi.org/10.1108/JKM-01-2020-0034

Pflaum, A.A., Gölzer, P., 2018. The IoT and digital transformation: Toward the data-driven enterprise. IEEE Pervasive Computing 17, 87–91. https://doi.org/10.1109/MPRV.2018.011591066

Pinsonneault, A., Kraemer, K.L., 2002. Exploring the role of information technology in organizational downsizing: a tale of two American cities. Organization Science 13, 191–208.

Pitafi, A.H., Rasheed, M.I., Kanwal, S., Ren, M., 2020. Employee agility and enterprise social media: The Role of IT proficiency and work expertise. Technology in Society 63, 101333. https://doi.org/10.1016/j.techsoc.2020.101333

Potdar, P.K., Routroy, S., Behera, A., 2017. Agile manufacturing: a systematic review of literature and implications for future research. Benchmarking: An International Journal 24, 2022–2048. https://doi.org/10.1108/BIJ-06-2016-0100

Rane, S.B., Narvel, Y.A.M., 2019. Re-designing the business organization using disruptive innovations based on blockchain-IoT integrated architecture for improving agility in future Industry 4.0. Benchmarking: An International Journal ahead-of-print. https://doi.org/10.1108/BIJ-12-2018-0445

Raut, R.D., Mangla, S.K., Narwane, V.S., Dora, M., Liu, M., 2021. Big Data Analytics as a mediator in Lean, Agile, Resilient, and Green (LARG) practices effects on sustainable supply chains. Transportation Research Part E: Logistics and Transportation Review 145, 102170. https://doi.org/10.1016/j.tre.2020.102170

Ravichandran, T., 2018. Exploring the relationships between IT competence, innovation capacity and organizational agility. The Journal of Strategic Information Systems 27, 22–42. https://doi.org/10.1016/j.jsis.2017.07.002

Rialti, R., Marzi, G., Silic, M., Ciappei, C., 2018. Ambidextrous organization and agility in big data era: The role of business process management systems. BPMJ 24, 1091–1109. https://doi.org/10.1108/BPMJ-07-2017-0210

Rialti, R., Zollo, L., Ferraris, A., Alon, I., 2019. Big data analytics capabilities and performance: Evidence from a moderated multi-mediation model. Technological Forecasting and Social Change 149, 119781. https://doi.org/10.1016/j.techfore.2019.119781

Saberi, S., Kouhizadeh, M., Sarkis, J., Shen, L., 2019. Blockchain technology and its relationships to sustainable supply chain management. International Journal of Production Research 57, 2117–2135. https://doi.org/10.1080/00207543.2018.1533261

Sambamurthy, Bharadwaj, Grover, 2003. Shaping Agility through Digital Options: Reconceptualizing the Role of Information Technology in Contemporary Firms. MIS Quarterly 27, 237. https://doi.org/10.2307/30036530

Saraf, N., Langdon, C.S., Gosain, S., 2007. IS Application Capabilities and Relational Value in Interfirm Partnerships. Information Systems Research 18, 320–339. https://doi.org/10.1287/isre.1070.0133

Sarker, Saonee, Munson, C.L., Sarker, Suprateek, Chakraborty, S., 2009. Assessing the relative contribution of the facets of agility to distributed systems development success: an Analytic Hierarchy Process approach. European Journal of Information Systems 18, 285–299. https://doi.org/10.1057/ejis.2009.25

Schwab, K., Zahidi, S., 2020. How Countries are Performing on the Road to Recovery, The Global Competitiveness Report. World Economic Forum, Davos.

Scuotto, V., Santoro, G., Bresciani, S., Giudice, M.D., 2017. Shifting intra- and inter-organizational innovation processes towards digital business: An empirical analysis of SMEs. Creativity and Innovation Management 26, 247–255. https://doi.org/10.1111/caim.12221

Seetharaman, P., 2020. Business models shifts: Impact of Covid-19. International Journal of Information Management 54, 102173. https://doi.org/10.1016/j.ijinfomgt.2020.102173

Sestino, A., Prete, M.I., Piper, L., Guido, G., 2020. Internet of Things and Big Data as enablers for business digitalization strategies. Technovation. https://doi.org/10.1016/j.technovation.2020.102173

Shams, R., Vrontis, D., Belyaeva, Z., Ferraris, A., Czinkota, M.R., 2021. Strategic agility in international business: A conceptual framework for "agile" multinationals. Journal of International Management 27, 100737. https://doi.org/10.1016/j.intman.2020.100737

Shams, R., Vrontis, D., Belyaeva, Z., Ferraris, A., Czinkota, M.R., 2020a. Strategic agility in international business: A conceptual framework for "agile" multinationals. Journal of International Management 100737. https://doi.org/10.1016/j.intman.2020.100737

Shams, R., Vrontis, D., Chaudhuri, R., Chavan, G., Czinkota, M.R., 2020b. Stakeholder engagement for innovation management and entrepreneurial development: A meta-analysis. Journal of Business Research 119, 67–86. https://doi.org/10.1016/j.jbusres.2020.08.036

Shashi, Centobelli, P., Cerchione, R., Ertz, M., 2020. Agile supply chain management: where did it come from and where will it go in the era of digital transformation? Industrial Marketing Management 90, 324–345. https://doi.org/10.1016/j.indmarman.2020.07.011

Sheel, A., Nath, V., 2019. Effect of blockchain technology adoption on supply chain adaptability, agility, alignment and performance. MRR 42, 1353–1374. https://doi.org/10.1108/MRR-12-2018-0490

Sherehiy, B., Karwowski, W., Layer, J.K., 2007. A review of enterprise agility: Concepts, frameworks, and attributes. International Journal of Industrial Ergonomics 37, 445–460. https://doi.org/10.1016/j.ergon.2007.01.007

Shiranifar, A., Rahmati, M., Jafari, F., 2019. Linking IT to supply chain agility: does knowledge management make a difference in SMEs? International Journal of Logistics Systems and Management 34, 123–138. https://doi.org/10.1504/IJLSM.2019.102066

Sjödin, D., Parida, V., Palmié, M., Wincent, J., 2021. How AI capabilities enable business model innovation: Scaling AI through co-evolutionary processes and feedback loops. Journal of Business Research 134, 574–587. https://doi.org/10.1016/j.jbusres.2021.05.009

Škare, M., Soriano, D.R., 2021. A dynamic panel study on digitalization and firm's agility: What drives agility in advanced economies 2009–2018. Technological Forecasting and Social Change 163, 120418. https://doi.org/10.1016/j.techfore.2020.120418

Sommer, A.F., 2019. Agile Transformation at LEGO Group: <i>Implementing Agile methods in multiple departments changed not only processes but also employees' behavior and mindset</i>. Research-Technology Management 62, 20–29. https://doi.org/10.1080/08956308.2019.1638486

Stylos, N., Zwiegelaar, J., Buhalis, D., 2021. Big data empowered agility for dynamic, volatile, and time-sensitive service industries: the case of tourism sector. IJCHM 33, 1015–1036. https://doi.org/10.1108/IJCHM-07-2020-0644

Swafford, P.M., Ghosh, S., Murthy, N., 2008. Achieving supply chain agility through IT integration and flexibility. International Journal of Production Economics 116, 288–297. https://doi.org/10.1016/j.ijpe.2008.09.002

Swafford, P.M., Ghosh, S., Murthy, N.N., 2006. A framework for assessing value chain agility. Int Jrnl of Op & Prod Mnagemnt 26, 118–140. https://doi.org/10.1108/01443570610641639

Tallon, P.P., 2008. Inside the adaptive enterprise: an information technology capabilities perspective on business process agility. Inf Technol Manage 9, 21–36. https://doi.org/10.1007/s10799-007-0024-8

Tallon, P.P., Pinsonneault, A., 2011. Competing Perspectives on the Link Between Strategic Information Technology Alignment and Organizational Agility: Insights from a Mediation Model. MIS Quarterly 35, 463–486. https://doi.org/10.2307/23044052

Tallon, P.P., Queiroz, M., Coltman, T., Sharma, R., 2019. Information technology and the search for organizational agility: A systematic review with future research possibilities. The Journal of Strategic Information Systems 28, 218–237. https://doi.org/10.1016/j.jsis.2018.12.002

Talwar, S., Kaur, P., Fosso Wamba, S., Dhir, A., 2021. Big Data in operations and supply chain management: a systematic literature review and future research agenda. International Journal of Production Research 59, 3509–3534. https://doi.org/10.1080/00207543.2020.1868599

Tandon, A., Dhir, A., Islam, N., Mäntymäki, M., 2020. Blockchain in healthcare: A systematic literature review, synthesizing framework and future research agenda. Computers in Industry 122, 103290.

Teece, D., Peteraf, M., Leih, S., 2016. Dynamic Capabilities and Organizational Agility: Risk, Uncertainty, and Strategy in the Innovation Economy. California Management Review 58, 13–35. https://doi.org/10.1525/cmr.2016.58.4.13

Teece, D.J., 2016. Dynamic capabilities and entrepreneurial management in large organizations: Toward a theory of the (entrepreneurial) firm. European Economic Review, The Economics of Entrepreneurship 86, 202–216. https://doi.org/10.1016/j.euroecorev.2015.11.006

Teece, D.J., 2007. Explicating dynamic capabilities: the nature and microfoundations of (sustainable) enterprise performance. Strategic management journal 28, 1319–1350.

Teece, D.J., Pisano, G., Shuen, A., 1997. Dynamic capabilities and strategic management. Strategic Management Journal 18, 509–533. https://doi.org/10.1002/(SICI)1097-0266(199708)18:7<509::AID-SMJ882>3.0.CO;2-Z

Todeschini, R., Baccini, A., 2016. Handbook of Bibliometric Indicators, Handbook of Bibliometric Indicators. Wiley-VCH Verlag GmbH & Co. KGaA, Weinheim, Germany. https://doi.org/10.1002/9783527681969

Tranfield, D., Denyer, D., Smart, P., 2003. Towards a Methodology for Developing Evidence-Informed Management Knowledge by Means of Systematic Review. British Journal of Management 14, 207–222. https://doi.org/10.1111/1467-8551.00375

Vagnoni, E., Khoddami, S., 2016. Designing competitivity activity model through the strategic agility approach in a turbulent environment. FS 18, 625–648. https://doi.org/10.1108/FS-03-2016-0012

Van Eck, N., Waltman, L., Van den Berg, J., Kaymak, U., 2006. Visualizing the computational intelligence field [Application Notes]. IEEE Computational Intelligence Magazine 1, 6–10. https://doi.org/10.1109/MCI.2006.329702

van Eck, N.J., Waltman, L., 2010. Software survey: VOSviewer, a computer program for bibliometric mapping. Scientometrics 84, 523–538. https://doi.org/10.1007/s11192-009-0146-3

Verhoef, P.C., Broekhuizen, T., Bart, Y., Bhattacharya, A., Qi Dong, J., Fabian, N., Haenlein, M., 2021. Digital transformation: A multidisciplinary reflection and research agenda. Journal of Business Research 122, 889–901. https://doi.org/10.1016/j.jbusres.2019.09.022

Vial, G., 2019. Understanding digital transformation: A review and a research agenda. The Journal of Strategic Information Systems, SI: Review issue 28, 118–144. https://doi.org/10.1016/j.jsis.2019.01.003

Vickery, S.K., Droge, C., Setia, P., Sambamurthy, V., 2010. Supply chain information technologies and organisational initiatives: complementary versus independent effects on agility and firm performance. International Journal of Production Research 48, 7025–7042. https://doi.org/10.1080/00207540903348353

Vokurka, R.J., Fliedner, G., 1998. The journey toward agility. Industrial Management & Data Systems 98, 165–171. https://doi.org/10.1108/02635579810219336

von Alberti-Alhtaybat, L., Al-Htaybat, K., Hutaibat, K., 2019. A knowledge management and sharing business model for dealing with disruption: The case of Aramex. Journal of Business Research 94, 400–407. https://doi.org/10.1016/j.jbusres.2017.11.037

von Rosing, M., Etzel, G., 2020. Introduction to the digital transformation lifecycle, in: CEUR Workshop Proceedings. pp. 92–99.

Walter, A.-T., 2021. Organizational agility: ill-defined and somewhat confusing? A systematic literature review and conceptualization. Manag Rev Q 71, 343–391. https://doi.org/10.1007/s11301-020-00186-6

Walter, A.T., 2020. Organizational agility: ill-defined and somewhat confusing? A systematic literature review and conceptualization. Management Review Quarterly 1–49. https://doi.org/10.1007/s11301-020-00186-6

Wang, G., Gunasekaran, A., Ngai, E.W.T., Papadopoulos, T., 2016. Big data analytics in logistics and supply chain management: Certain investigations for research and applications. International Journal of Production Economics 176, 98–110. https://doi.org/10.1016/j.ijpe.2016.03.014

Wang, Y., Han, J.H., Beynon-Davies, P., 2019. Understanding blockchain technology for future supply chains: a systematic literature review and research agenda. Supply Chain Management: An International Journal 24, 62–84. https://doi.org/10.1108/SCM-03-2018-0148

Wang, Z., Pan, S.L., Ouyang, T.H., Chou, T.-C., 2013. Achieving IT-enabled enterprise agility in China: an IT organizational identity perspective. IEEE Transactions on Engineering Management 61, 182–195. https://doi.org/10.1109/TEM.2013.2259494

Warner, K.S.R., Wäger, M., 2019. Building dynamic capabilities for digital transformation: An ongoing process of strategic renewal. Long Range Planning 52, 326–349. https://doi.org/10.1016/j.lrp.2018.12.001

Webster, J., Watson, R.T., 2002. Analyzing the past to prepare for the future: Writing a literature review. MIS quarterly xiii–xxiii.

Westerman, G., Calméjane, C., Bonnet, D., Ferraris, P., McAfee, A., 2011. Digital Transformation: a roadmap for billion dollar organizations. Capgemini Consulting & MIT Center for Digital Business, Paris & Cambridge.

White, A., Daniel, E.M., Mohdzain, M., 2005. The role of emergent information technologies and systems in enabling supply chain agility. International Journal of Information Management 25, 396–410. https://doi.org/10.1016/j.ijinfomgt.2005.06.009

Wixom, B.H., Watson, H.J., 2001. An empirical investigation of the factors affecting data warehousing success. MIS quarterly 17–41.

Yusuf, Y.Y., Sarhadi, M., Gunasekaran, A., 1999. Agile manufacturing:: The drivers, concepts and attributes. International Journal of Production Economics 62, 33–43. https://doi.org/10.1016/S0925-5273(98)00219-9

Zhang, Z., Sharifi, H., 2000. A methodology for achieving agility in manufacturing organisations. International Journal of Operations & Production Management 20, 496–513. https://doi.org/10.1108/01443570010314818

Zhen, J., Cao, C., Qiu, H., Xie, Z., 2021. Impact of organizational inertia on organizational agility: the role of IT ambidexterity. Information Technology and Management 22, 53–65.

Zhou, J., Mavondo, F.T., Saunders, S.G., 2019. The relationship between marketing agility and financial performance under different levels of market turbulence. Industrial Marketing Management 83, 31–41. https://doi.org/10.1016/j.indmarman.2018.11.008

Zielske, M., Held, T., 2020. The Use of Agile Methods in Logistics Start-ups: An Explorative Multiple Case Study. International Journal of Innovation and Technology Management (IJITM) 17, 1–24.

Zupic, I., Čater, T., 2015. Bibliometric Methods in Management and Organization. Organizational Research Methods 18, 429–472. https://doi.org/10.1177/1094428114562629

**Author Biographic Notes**

**Francesco Ciampi** is a Full Professor of Management at the Department of Economics and Management of the University of Florence. His research is mainly focused on innovation management, knowledge management, company strategy, organizational ambidexterity, default prediction modelling, management consulting, and small and medium-sized enterprises. He has authored and co-authored a number of papers that appeared in conferences, edited books and journals such as Small Business Economics, Scientometrics, Journal of Business Research, International Journal of Information Management, IEEE Transactions on Engineering Management, The Journal of Technology Transfer, Technological Forecasting and Social Change, Journal of Small Business Management, Management Decision, Journal of Knowledge Management, Journal of Intellectual Capital. He is currently an Associate Editor of the Journal of Intellectual Capital and member of the Board of Directors of the Italian Management Society. He has been teaching under-graduate courses in Management since 1995 and post-graduate courses in Management Consulting since 2004.

**Monica Faraoni** is an Associate Professor of Management at the University of Florence. She received a PhD in management from the University of Bologna and was visiting professor at the Wharton Business School of the University of Pennsylvania, Philadelphia. Her research interests center around how new technologies and associated digitalization of information impact consumer and firm behavior with particular attention to big data, online consumer purchasing process and brand management in fashion industry. She is the author of numerous publications in international journals such as Journal of Business Research, Journal of Interactive Marketing, Journal of Knowledge Management.

**Jacopo Ballerini** is a lecturer of Innovation Management and Digital Transformation at the University of Turin and a PhD candidate in Business and Management. He has been a marketing professional with an early career started at the age of 24 and developed through years of steady managerial experiences in Multinationals and Unicorn Start-ups in the Consumer goods industry, especially in DYI and Consumer Electronics, as Product Marketing Manager, and after, as Key Account Manager. After a proven record of success in managing different product categories through both traditional offline and digital channels, his professional career has led him to develop a particular scientific interest in the dynamics of digital marketing, online sales and social media.

**Francesco Meli** is a scholarship holder at the Department of Economics and Management of the University of Florence. He holds an MSc in Management from the University of Florence. His research is mainly focused on Big Data, Collaborative Innovation, Co-creation, Knowledge Management and New Product Development, especially using Bibliometric Analysis and Systematic Literature Review methodologies.